\documentclass[amsmath,amssymb,amsbsy,prl,twocolumn,tightenlines,superscriptaddress,floatfix]{revtex4-1}
\usepackage{amsfonts} % collection of miscellaneous TeX fonts
\usepackage{amsmath} 
\usepackage{amssymb}
\usepackage{physics}
\usepackage{graphicx}
\usepackage{nicefrac}
\usepackage{blindtext}
\usepackage{lineno}
\usepackage{cancel}
\usepackage{scalerel}
\usepackage[normalem]{ulem}
\usepackage{xfrac}
\usepackage{bm}% bold math
\usepackage[usenames, dvipsnames]{color} % use colored tex
\usepackage{dcolumn}
\usepackage{epic} % enhanced picture environment
\usepackage{epsfig}
\usepackage{eucal} % changes the font selected by \mathcal
\usepackage{esdiff}
\usepackage{graphicx}
\usepackage[breaklinks,colorlinks = true,linkcolor = magenta,urlcolor=magenta,citecolor=red]{hyperref}
\usepackage{mathrsfs}%\mathscr
\usepackage{soul}
\sethlcolor{yellow} % Define the highlight color
\usepackage{subfigure}
\usepackage{textcomp}
\usepackage{times}
\usepackage{verbatim}
\usepackage{wrapfig} % include figure with text wrapping around
\usepackage{xy} % package for drawing complex diagrams
\usepackage{lineno}
\usepackage{float}
\usepackage{import}
\newcommand{\dop}[1]{\qty[\dv{}{t}+{#1}]}
\newcommand{\Pb}[1]{{\mathbb{P}}^{{({#1})}}}
\newcommand{\Pc}[1]{{\mathcal{P}}^{{({#1})}}}
\newcommand{\Cc}[1]{{\mathcal{C}}^{{({#1})}}}
\newcommand{\Iq}{\displaystyle \int_{}^{}\dd{\vb{q}}}
\newcommand{\Ip}{\displaystyle \int_{}^{}\dd{\vb{p}}}
\newcommand{\Ir}{\displaystyle \int_{}^{}\dd{\vb{r}}}
\newcommand{\DqDp}{\dd{q}\;\dd{p}}
\newcommand{\Itr}{\displaystyle \int_{{\bm{\Delta} _{{\vb{k}}}}}}
\newcommand{\Iptr}{\displaystyle \int_{{\Delta _{{k}}}}}
\newcommand{\Trd}{{\Delta _{{k}}}}

\newcommand{\gc}[1]{{{#1}\mathstrut}}
\newcommand{\gcr}[1]{{{#1}\mathstrut}^{{(3)\mathstrut}}}
\newcommand{\gcw}[1]{{{#1}\mathstrut}^{{(2)\mathstrut}}}
\newcommand{\zf}[1]{{\mathcal{S}}^{{{#1}\mathstrut}}}
\newcommand{\uf}[1]{{\mathcal{U}}^{{{#1}\mathstrut}}}
\newcommand{\tf}[1]{{\mathcal{T}}^{{{#1}\mathstrut}}}

\newcommand{\Btrf}[2]{{\mathrm{T}}^{{{#2}\mathstrut}}_{{{#1}\mathstrut}}}
\newcommand{\tn}[2]{{\mathrm{T}}_{{{#2}\mathstrut}}^{{{#1}\mathstrut}}}
\newcommand{\tnf}[1]{{\mathrm{T}}^{{{#1}\mathstrut}}}
\newcommand{\ef}[1]{{E}^{{{#1}\mathstrut}}}

\newcommand{\Oth}[1]{\hat{\bm{\theta }}\qty(\omega_{{#1}}^u)}
\newcommand{\OthB}[1]{\hat{\bm{\theta }}\qty(\omega_{{#1}}^b)}

\newcommand{\edq}{EDQNM }
\newcommand{\mhd}{MHD }
\newcommand{\edqmhd}{EDQNM-MHD }
\newcommand{\dL}{d_{\mathrm{L}}}
\newcommand{\dU}{d_{\mathrm{U}}}
\newcommand{\pde}{PDE }

\begin{document}
%HHHHHHHHHHHHHHHHHHHHHHHHHHHHHHHHHHHHHHHHHHHHHHHHHHHHHHHHHHHHHHHHHHHHHHHHHHHHHH
\title{An upper critical dimension for dynamo action: A $d$-dimensional closure model study}
\author{Sugan Durai Murugan} 
\email{vsdmfriend@gmail.com}
%\affiliation{International Centre for Theoretical Sciences, Tata Institute of Fundamental Research, Bengaluru 560089, India} 
\affiliation{Department of Mechanical Engineering, Johns Hopkins University, Baltimore, Maryland 21218, USA} 
\author{Giorgio Krstulovic}
\email{krstulovic@oca.eu} 
\affiliation{Universit\'e C\^ote d'Azur, Observatoire de la C\^ote d'Azur, CNRS, Laboratoire Lagrange, Boulevard de l'Observatoire CS 34229—F 06304 Nice Cedex 4, France}
 \author{Dario Vincenzi} 
 \email{dario.vincenzi@univ-cotedazur.fr}
 \email{Also Associate, International Centre for Theoretical Sciences, Tata Institute of Fundamental Research, 
 Bengaluru 560089, India} 
 \affiliation{Universit\'e C\^ote d'Azur, CNRS, LJAD, 06100 Nice, France} 
\author{Samriddhi Sankar Ray} 
\email{samriddhisankarray@gmail.com}
\affiliation{International Centre for Theoretical Sciences, Tata Institute of Fundamental Research, Bengaluru 560089, India}
%HHHHHHHHHHHHHHHHHHHHHHHHHHHHHHHHHHHHHHHHHHHHHHHHHHHHHHHHHHHHHHHHHHHHHHHHHHHHHH
\begin{abstract} 

We construct a $d$-dimensional Eddy Damped Quasi-Normal Markovian (EDQNM)
	Closure Model to study dynamo action in arbitrary dimensions.  In
	particular, we find lower $\dL$ and upper $\dU$ critical dimensions for sustained dynamo action in this
	incompressible problem. Our model is adaptable for future studies incorporating helicity,
	compressible effects and a wide range of magnetic Reynolds and Prandtl numbers.

\end{abstract}
%HHHHHHHHHHHHHHHHHHHHHHHHHHHHHHHHHHHHHHHHHHHHHHHHHHHHHHHHHHHHHHHHHHHHHHHHHHHHHH
\date{\today}
\maketitle
%HHHHHHHHHHHHHHHHHHHHHHHHHHHHHHHHHHHHHHHHHHHHHHHHHHHHHHHHHHHHHHHHHHHHHHHHHHHHHH

Large  magnetic  fields  are  at  the  heart  of  almost  every  observation  in
astrophysics; indeed, they play  a  pivotal  role  in,  as  well  as  shape  the
consequence of, the dynamics of  phenomena  ranging  from  star  formation,  the
interstellar    medium     to     the     underpinnings     of     the     solar
wind~\cite{ogilvie_astrophysical_2016,                brandenburg_advances_2018,
brandenburg_galactic_2023}. And yet questions remain how such sustained magnetic
fields    arise---the    \emph{dynamo}    problem---in    the    first     place
\cite{moffatt_selfexciting_2019,   rincon_dynamo_2019,   tobias_turbulent_2021}.
Since astrophysical flows are also,  typically,  notoriously  turbulent,  it  is
natural  to  look  for  answers  to  such  questions  within  the  framework  of
magnetohydrodynamic        (MHD)        turbulence~\cite{choudhuri_physics_1998,
davidson_introduction_2001,                    biskamp_magnetohydrodynamic_2003,
galtier_introduction_2016, schekochihin_mhd_2022}. While a theory for the dynamo
problem rooted in the full set of equations for  MHD  is  desirable,  there  are
formidable challenges to this. From  the  point  of  view  of  direct  numerical
simulations (DNSs) of such systems, the parameter  space  accessible  to  modern
simulations are quite far  from  what  is  realisable  in  either  astrophysical
systems   or   liquid-metal   experiments~\cite{monchaux_generation_2007}.   For
example, the Prandtl number, defined as the ratio of the  kinetic  viscosity  to
the magnetic diffusivity ${\rm Pm} \equiv \nu/\eta$, range from values as  large
as $10^{14}$ (interstellar medium) to those as small as $10^{-5}$ (liquid sodium
experiments). Such a range of numbers are prohibitively expensive for DNSs; thus
more often than not, theoretical  approaches  based  on  reasonable  assumptions
provide additional insights and a fresh perspective in understanding the nuances
of the dynamo problem.

An excellent example of such theoretical approaches, and the deep insights  they
provide,   is   the    Kazantsev    model    for    the    fluctuation    dynamo
\cite{kazantsev_enhancement_1968}. In this stochastic model, the velocity  field
is Gaussian and statistically homogenous, isotropic, and  parity  invariant.  In
addition, the correlation time is assumed to be  zero---probably  the  strongest
simplification  in  this  model.  By  varying  the  features  of   the   spatial
correlations of the velocity field, it is possible to study the magnetic  growth
as a function of  the  degree  of  compressibility  of  the  flow,  its  spatial
regularity, the space dimension, and the Prandtl and magnetic  Reynolds  numbers
(see,            \textit{e.g.},            Refs.~\cite{falkovich_particles_2001,
brandenburg_astrophysical_2005,brandenburg_advances_2018,    rincon_dynamo_2019,
tobias_turbulent_2021}). In particular, the Kazantsev  model  has  provided  the
first evidence of the existence of a maximum critical dimension for  the  dynamo
effect beyond which this random flow becomes unable to amplify a magnetic  field
\cite{gruzinov_small-scale-field_1996, schekochihin_spectra_2002}. The range  of
dimensions where there is dynamo shrinks as the velocity becomes less  and  less
regular in space, until it vanishes when the H\"older exponent of  the  velocity
falls below 1/2 \cite{arponen_dynamo_2007}. Compressibility,  however,  has  the
effect of widening the range of dimensions over which the dynamo is possible  in
this model \cite{martins_afonso_kazantsev_2019}. Interestingly, dimension  $d  =
3$ is the one where the least flow regularity is required for the dynamo  effect
to take place, independently of the degree of compressibility.
 
It is easy to appreciate why theoretical models with variable roughness (of  the
velocity field) and compressibility  have  a  direct  bearing  on  understanding
\emph{real} dynamos. Nevertheless and especially given the strong  parallels  of
this problem to critical phenomena and phase transitions, the role of dimensions
in the dynamo---no-dynamo transition deserves some attention. Taking this  point
of  view  and  recalling  the  fundamental  discoveries---such  as   dimensional
regularization or the $4-\epsilon$  expansion~\cite{wilson_critical_1972}---made
possible by going beyond the physically obvious $d=2$ or $3$ dimensions,  it  is
not unreasonable  to  ask  if  there  is  an  analogue  of  a  \emph{lower}  and
\emph{upper critical dimension} below and  beyond  which,  respectively,  dynamo
action ceases to be. Indeed, such a point of view, of  going  beyond  physically
realisable integer dimensions of two and three, has lead to interesting  results
on    intermittency    and    energy     cascades     in     classical     fluid
turbulence~\cite{fournier_d-dimensional_1978, lvov_quasi-gaussian_2002,
celani_turbulence_2010, frisch_turbulence_2012, ray_thermalized_2015, ray_non-intermittent_2018, picardo_lagrangian_2020}.  In
this  paper,  we  simply  ask  if  there  are  lower  $\dL$  and   upper   $\dU$
\emph{critical} dimension within which dynamo action is confined?

While it is desirable to overcome the assumptions of  Gaussianity  and  temporal
decorrelation of the Kazantsev model and at the same  time  consider  the  fully
nonlinear regime, it is difficult to answer the above question through  DNSs  in
arbitrary dimension $d$. Instead, we construct a $d$-dimensional  closure  model
for MHD turbulence, which in the absence of a magnetic  field,  reduces  to  the
well-known   Eddy-Damped    Quasi-Normal    Markovian    (EDQNM)    for    fluid
turbulence~\cite{kraichnan_inertial_1967,                orszag_analytical_1970,
fournier_d-dimensional_1978,                                 rose_ha_fully_1978,
clark_effect_2021,clark_critical_2022}.  We  then  perform  detailed   numerical
simulations to  show  that  for  a  given  magnetic  (\textrm{Rm})  and  kinetic
(\textrm{Re}) Reynolds number the dynamo action is  constrained  for  dimensions
$\dL \leqslant d  \leqslant  \dU$,  with  the  lower  critical  dimension  $\dL$
marginally higher than 2 and a finite  upper  critical  dimension  $\dU$  beyond
which the dynamo cannot be sustained.

The first question is of course how do we construct this $d$-dimensional closure
model for MHD turbulence? Theoretically, the full MHD equations suffer from  the
same closure problems---and hence analytical  progress---as  the  Navier-Stokes
equation for fluid turbulence~\cite{lesieur_turbulence_2008}. We recall that  in
fluid  turbulence,  theoretical  progress   in   understanding   the   two-point
correlation function stems first from a Quasi-Normal approximation which  allows
rewriting fourth-order moments as sums of  products  of  different  second-order
moments. Then, the successive use of an (phenomenological) eddy-damping rate and
\emph{Markovianization} leads to a closed equation for the fluid kinetic  energy
spectrum in the \edq model. We follow a similar  approach,  beginning  with  the
incompressible MHD equations, to derive  the  corresponding  equations  for  the
fluid $\ef{u}(k)$ and magnetic $\ef{b}(k)$ energy spectra:

\begin{subequations}
\begin{align}
\dv{}{t}\ef{u}(k,t)&= \tn{u}{(s)}(k)+\tn{u}{(c)}(k) - 2\nu k^2 \ef{u}(k,t); \label{eq:edqnm-mhd_spectrum_u}\\
\dv{}{t}\ef{b}(k,t)&= \tn{b}{(s)}(k)+\tn{b}{(c)}(k) - 2\eta k^2 \ef{b}(k,t). \label{eq:edqnm-mhd_spectrum_b}
\end{align}%
\label{eq_main:edqnm-mhd_spectrum}%
\end{subequations}%
The transfer terms  are conveniently expressed in a form which underlines the distinct contributions 
from the self [subscript (s)] and coupled [subscript (c)] terms:
\begin{widetext}
\begin{subequations}
\begin{align}
\tn{u}{(s)}(k)&= 8K_d \Iptr \DqDp W_d\qty(\Trd) \theta^u_{kpq}\frac{k}{pq}\gc{b}_{kpq}\qty[k^{d-1}\ef{u}(p)-p^{d-1}\ef{u}(k)]\ef{u}(q); \label{eq:transfer_u_s}\\
\tn{u}{(c)}(k)&= 8K_d \Iptr \DqDp W_d\qty(\Trd) \theta^b_{kpq}\frac{k}{pq}\gc{c}_{kpq}\qty[k^{d-1}\ef{b}(p)-p^{d-1}\ef{u}(k)]\ef{b}(q); \label{eq:transfer_u_c}\\
\tn{b}{(s)}(k)&= 8K_d \Iptr \DqDp W_d\qty(\Trd) \theta^b_{qkp}\frac{k}{pq}\gc{h}_{kpq}\qty[k^{d-1}\ef{b}(p)-p^{d-1}\ef{b}(k)]\ef{u}(q); \label{eq:transfer_B_s}\\
\tn{b}{(c)}(k)&= 8K_d \Iptr \DqDp W_d\qty(\Trd) \theta^b_{pqk}\frac{p}{kq}\gc{c}_{pkq}\qty[k^{d-1}\ef{u}(p)-p^{d-1}\ef{b}(k)]\ef{b}(q). \label{eq:transfer_B_c}
\end{align}
\label{eq_main:transfer_split2}
\end{subequations}
\end{widetext}
In Eqs.~\eqref{eq_main:edqnm-mhd_spectrum}-\eqref{eq_main:transfer_split2},  and
what follows, $k$, $p$, and $q$ are wavenumbers and the superscripts  $(u)$  and
$(b)$ always denote the fluid and magnetic fields, respectively.  The  integrals
are over triads $\Trd$ formed from triangles with sides ${\bf  k}$,  ${\bf  p}$,
${\bf q}$, and the time-scales $\theta_{kpq}^{u}$ and $\theta_{kpq}^{b}$  are  a
consequence of the  eddy-damping  and  Markovian  assumption.  Furthermore,  the
explicit role of dimensions, which arise from the geometry of  these  triads  in
$d$-dimensional space, lead to an explicitly dimensional  prefactor  $K_d$,  the
weight of different triadic contributions $W_d(\Trd)$, the coupling coefficients
$\gc{b}_{kpq}$, $\gc{c}_{kpq}$, $\gc{h}_{kpq}$ and $\gc{c}_{pkq}$.
  We refer the reader to Appendices A-C for a full and  complete  derivation  of
  these equations as well as the precise form of each of the
terms and prefactors.

the basic phenomenology of the primitive mhd equations are already  apparent  in
the structure of our closure  model.  the  self-interaction  terms  ensures  the
transfer of energy from  different  wavenumbers  while  $\int_{0}^{\infty}\dd{k}
\tn{u}{(s)}(k)=   \int_{0}^{\infty}\dd{k}   \tn{b}{(s)}(k)=0$    ensuring    the
conservation of energy.  further,  the  cross  or  coupling  terms  mediate  the
transfer of energy between the fluid and magnetic fields and, again for  reasons
of  energy   conservation,   obey   $\int_{0}^{\infty}\dd{k}   \qty(\tn{u}{(c)}+
\tn{b}{(c)})=0$. finally, it is easy to check  that,  for  zero  magnetic  field
$\ef{b}  =  0$,  our  model  reduces  to   the   $d$-dimensional   fluid   edqnm
equations~\cite{clark_effect_2021}; similarly for $\ef{b} \neq 0$  and  choosing
$d = 2$ or $d =  3$,  we  recover  the  two  or  three-dimensional  \edq  model,
respectively,      for    MHD      turbulence      ~\cite{pouquet_strong_1976,
schilling_triadic_2002, pouquet_two-dimensional_1978}.

Trivially the dynamo question hinges on whether or not the total magnetic energy
$\mathcal{E}^b = \int_{0}^{\infty}\dd{k} \ef{b}(k)$ grows in time and eventually
saturates to a  nonzero  value  in  the  nonlinear  regime.  Starting  from  the
evolution equations, it is easy  to  show  that  for  an  initial  ($t=0$)  seed
magnetic field such  that  the  initial  energies  follow  $\mathcal{E}^u_0  \gg
\mathcal{E}^b_0$ (allowing for terms quadratic in  the  magnetic  energy  to  be
omitted)

\begin{align}
	\dv{\mathcal{E}^b}{t} \approx \int_{0}^{\infty}\dd{k} \qty(8K_d\lambda (k)-2\eta k^2) \ef{b}(k)
  \label{eq:dynamo_rate_spectral}
\end{align}
with  $\lambda_d  (k)   =   \Iptr   \DqDp   W_d(\Trd)   \theta_{pkq}\frac{p}{kq}
\gc{c}_{pqk} q^{d-1} \ef{u}(p)$ depending only on the properties  of  the  fluid
and the dimension $d$. Equation~\eqref{eq:dynamo_rate_spectral} shows  that  the
time behavior of $\mathcal{E}^b$ is the result
  of two opposing effects, namely magnetic diffusion and  the  amplification  by
  the velocity field,  and  it  depends  crucially  on  how  kinetic  energy  is
  distributed across the Fourier modes of the velocity.

%Assuming an exponential form $\mathcal{E}^b(t)\sim \exp(\overline{\lambda} t)$, it follows
%\begin{align}
%  \overline{\lambda }(t)&= \dfrac{\displaystyle \int_{0}^{\infty}\dd{k} \qty(\lambda (k)-2\eta k^2) \ef{b}(k)}{\displaystyle \int_{0}^{\infty}\dd{k} \ef{b}(k)}
%  \label{eq:dynamo_rate}
%\end{align}
%with the inevitable conclusion that for dynamo action $\overline{\lambda}$ should be non-negative for all time.

For ideal ($\nu = \eta = 0$) MHD, and with a finite number of  modes  (intrinsic
to the MHD-EDQNM model), further progress is  possible  theoretically.  This  is
because for $d > 2$, a global equipartition emerges as a thermal fixed point  of
the model. This ensures that for all dimensions $d>2$, the seed  magnetic  field
grows to asymptotically  (in  time)  reach  a  state  with  $\ef{u}(k)=\ef{b}(k)\sim k^{d-1}$. 
However, $d  =
2$  is  special.  Here,  the  conservation  of  magnetic  potential   ($\sum
\ef{b}(k)/k^2 $), an additional constraint on these modes ensures  a  lack
of global equipartition and thence $\mathcal{E}^u  \gg  \mathcal{E}^b$  for  all
times as long as $\mathcal{E}^u_0 \gg \mathcal{E}^b_0$. Thus, in the  ideal  and
finite-dimensional model, dynamo action is strictly possible for all  dimensions
$d > 2$.

But what happens for real flows which are  dissipative  and  out-of-equilibirum?
Here things are much harder to assess theoretically and we  resort  to  evidence
from numerical simulations to guide our intuition. We perform detailed numerical
simulation            of             our             MHD-EDQNM             model
(Eqs.~\eqref{eq_main:edqnm-mhd_spectrum}--\eqref{eq_main:transfer_split2})    in
dimensions $2\le d \le 12$, with $\nu = \eta =  5\times  10^{-4}$,  use  minimum
$k_{\rm min} = 2^{-3}$ and maximum $k_{\rm max} =  2^{10}$  wavenumbers,  and  a
time-stepping  $\delta  t  =  2\times  10^{-5}$,  allowing  us   to   obtain   a
well-resolved inertial range. Further details on the  numerical  set-up  and  in
particular how the wavenumbers are discretised are given in Appendix D.

We set up the numerical study of the dynamo problem in the following fashion.
We first develop a statistically stationary state for the kinetic spectrum by
keeping the magnetic field switched off and driving the kinetic energy spectrum
through a forcing spectrum $F(k)$ concentrated at large scales via $F(k)\sim
k^2 \exp(-k^2/2k_{I}^{2})$, with $k_I$ setting the injection scale. This
injection of energy is balanced with the net viscous dissipation rate
$\epsilon^u = \int_{0}^{\infty}\dd{k} 2\nu k^2 \ef{u}(k)$ to ensure a constant
net kinetic energy $\mathcal{E}^u \equiv \int_{0}^{\infty}\dd{k}
\ef{u}(k)=1.0$.

The  steady  state  is
characterized by  a  Kolmogorov  spectrum  $\ef{u}(k)\sim  k^{-5/3}$  for
$d\gtrapprox 3$  (albeit  with  an  ever-pronounced  bottleneck  effect  as  $d$
increases~\cite{clark_effect_2021}) or a $k^{-3}$ spectrum (due to  the  inverse
cascade)    for    $d\approx    2$     cases     ~\cite{kraichnan_inertial_1967,
frisch_crossover_1976,  fournier_d-dimensional_1978}. The  magnetic   spectrum
interaction is  switched  on  with  a  initial  seed  of  magnetic  energy  with
$\mathcal{E}^b_0=10^{-2}$. With the interaction  on,  the  forcing  is  adjusted
slightly to match the net dissipation  rate  which  now  includes  the  magnetic
dissipation rate $\epsilon^b$. In what follows, the time when the magnetic field
is switched on is set as  $t  =  0$,  and  the  dynamo  problem  is  studied  at
subsequent times.

\begin{figure}[t!]\centering
\includegraphics[width=1.0\columnwidth]{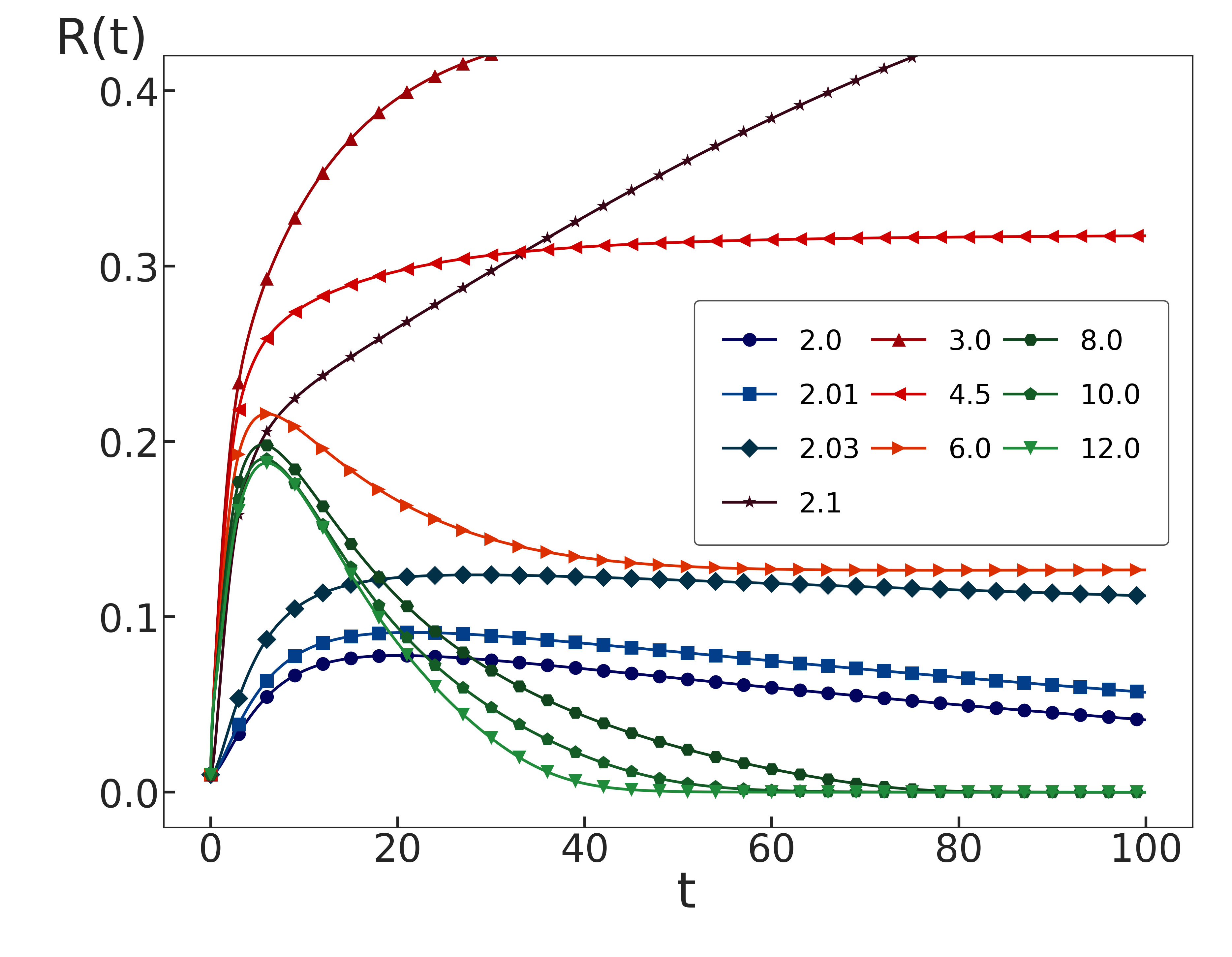}
\caption{A plot of $R(t)$ vs $t$ for several different
  dimensions. For dimensions $d \lesssim \dL \approx 2.04$ and  $d  \gtrsim  \dU
  \approx 6.5$, the magnetic energy,
	after an increasing initially, decreases  with  time  indicating  no  sustained
	dynamo action. For dimensions $\dL  \lesssim  d  \lesssim  \dU$,  the  magnetic
	energy increases in time with an eventual dimension-dependent saturation.}
\label{fig:dynamo_growth}
\end{figure}

To study the dynamo effect, we find it  useful  to  define  the  measure  $R(t)=
\dfrac{\mathcal{E}^b(t)}{\mathcal{E}^u(t)}$ and observe its  temporal  behaviour
for   different   dimensions.   In   Fig.~\ref{fig:dynamo_growth},    we    show
representative plots of $R(t)$ versus time for several different dimensions. For
two   dimensional   flows   and    as    expected~\cite{zeldovich_magnetic_1957,
zeldovich_magnetic_1980_jetp}, we have no dynamo  action  as  $R(t)$,  after  an
initial growth, decays in time. The three-dimensional case  is  just  as  clear:
$R(t)$ increases and eventually saturates to value slightly larger than 0.5 (not
shown) indicating dynamo action. What is interesting is the behaviour for  other
dimensions. Clearly, there  seems  to  be  dimensions  $d\gtrsim  2.0$  as  well
dimensions much larger than $d = 3.0$ where the dynamo fails. In fact in  higher
dimensions we do see an initial rise in $R(t)$ that becomes  unsustainable  with
time. All of this suggests that at  least  within  the  MHD-EDQNM  phenomenology
there must exist a lower critical dimension $\dL \gtrsim 2.0$ and a finite upper
critical dimension $\dU \gg 3.0$ which  dictates  the  dynamo---no-dynamo  phase
boundary.

\begin{figure*}[t!]\centering
\includegraphics[width = 0.33\textwidth]{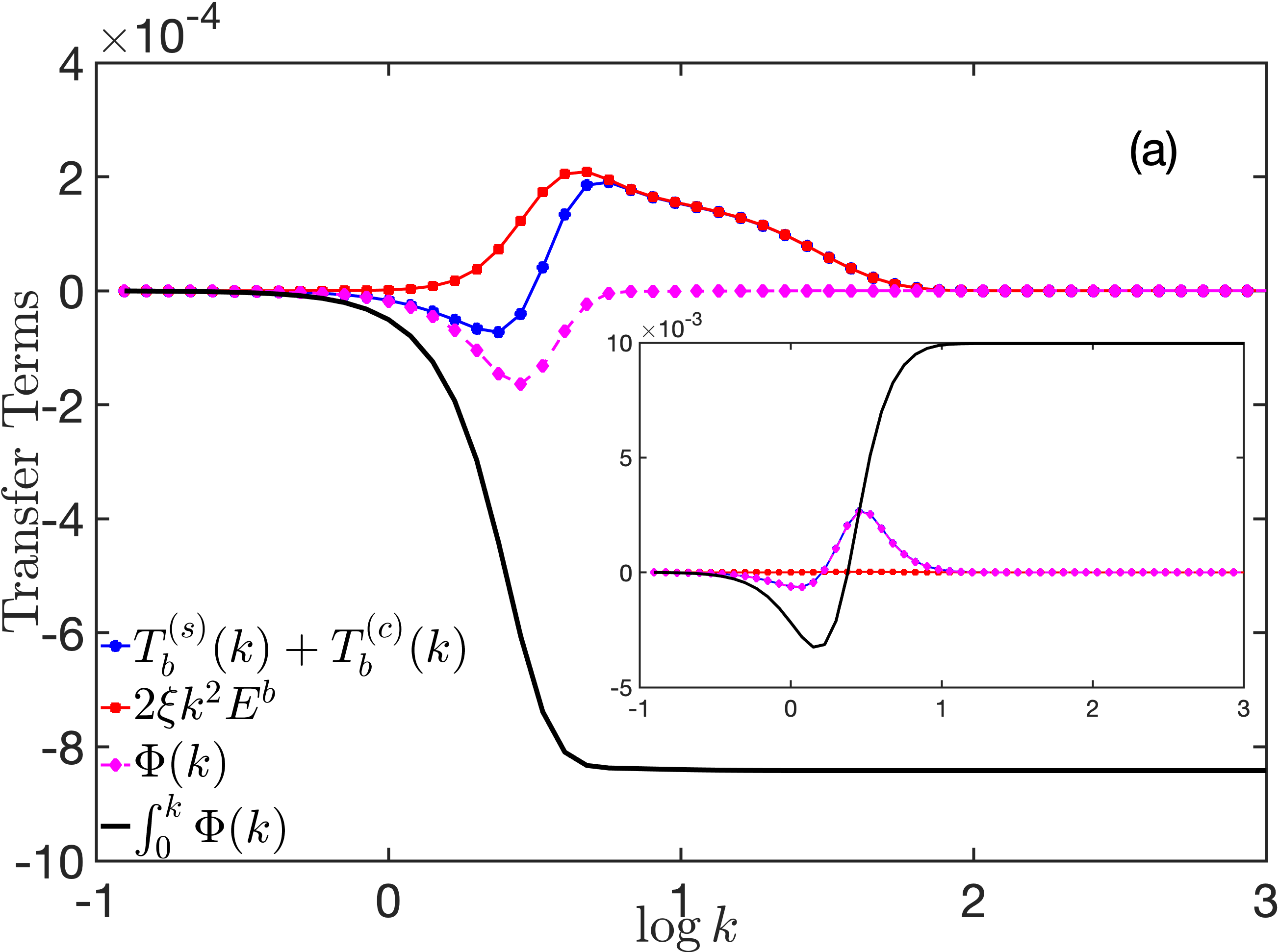}
\includegraphics[width = 0.33\textwidth]{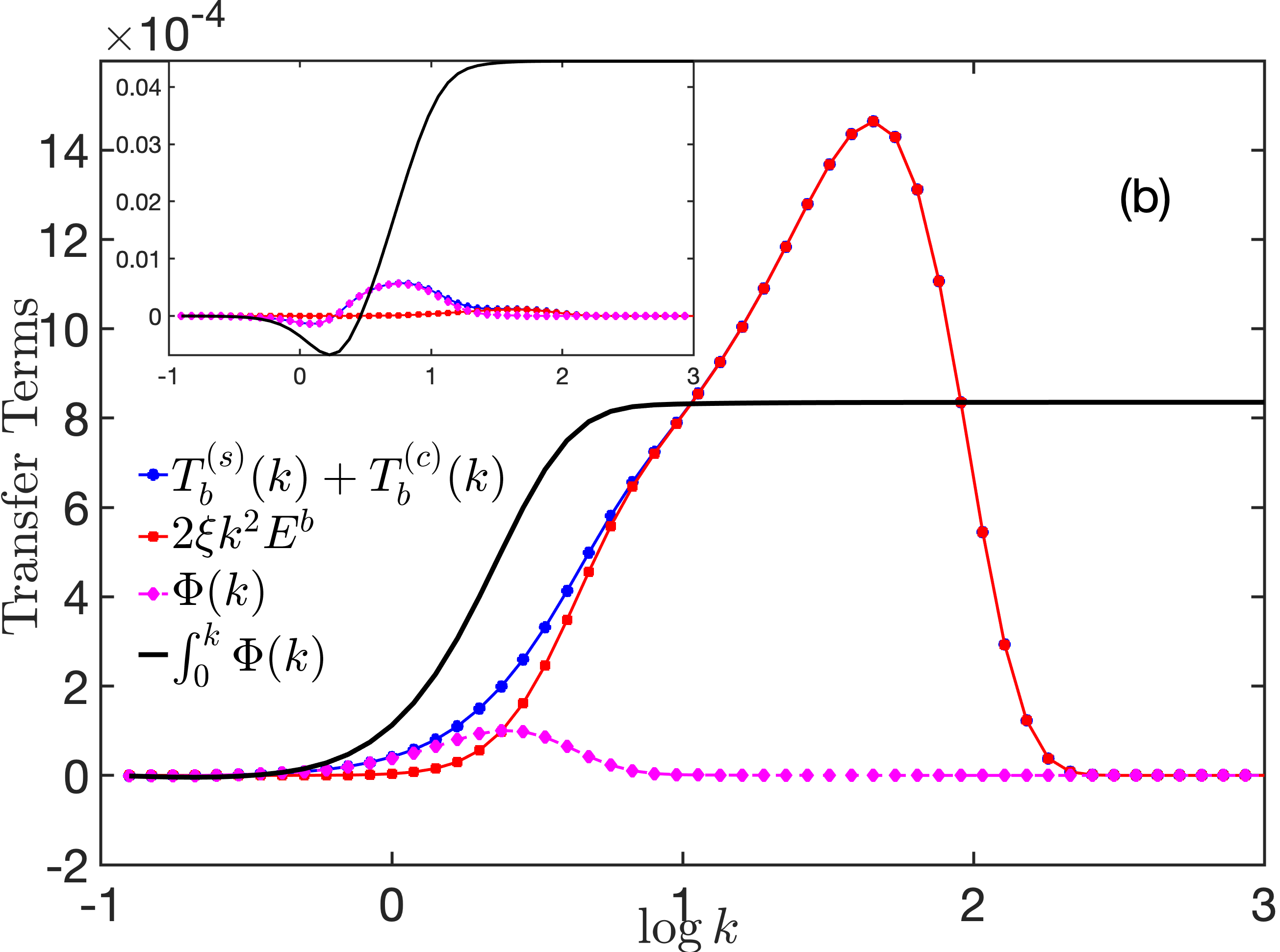}
\includegraphics[width = 0.33\textwidth]{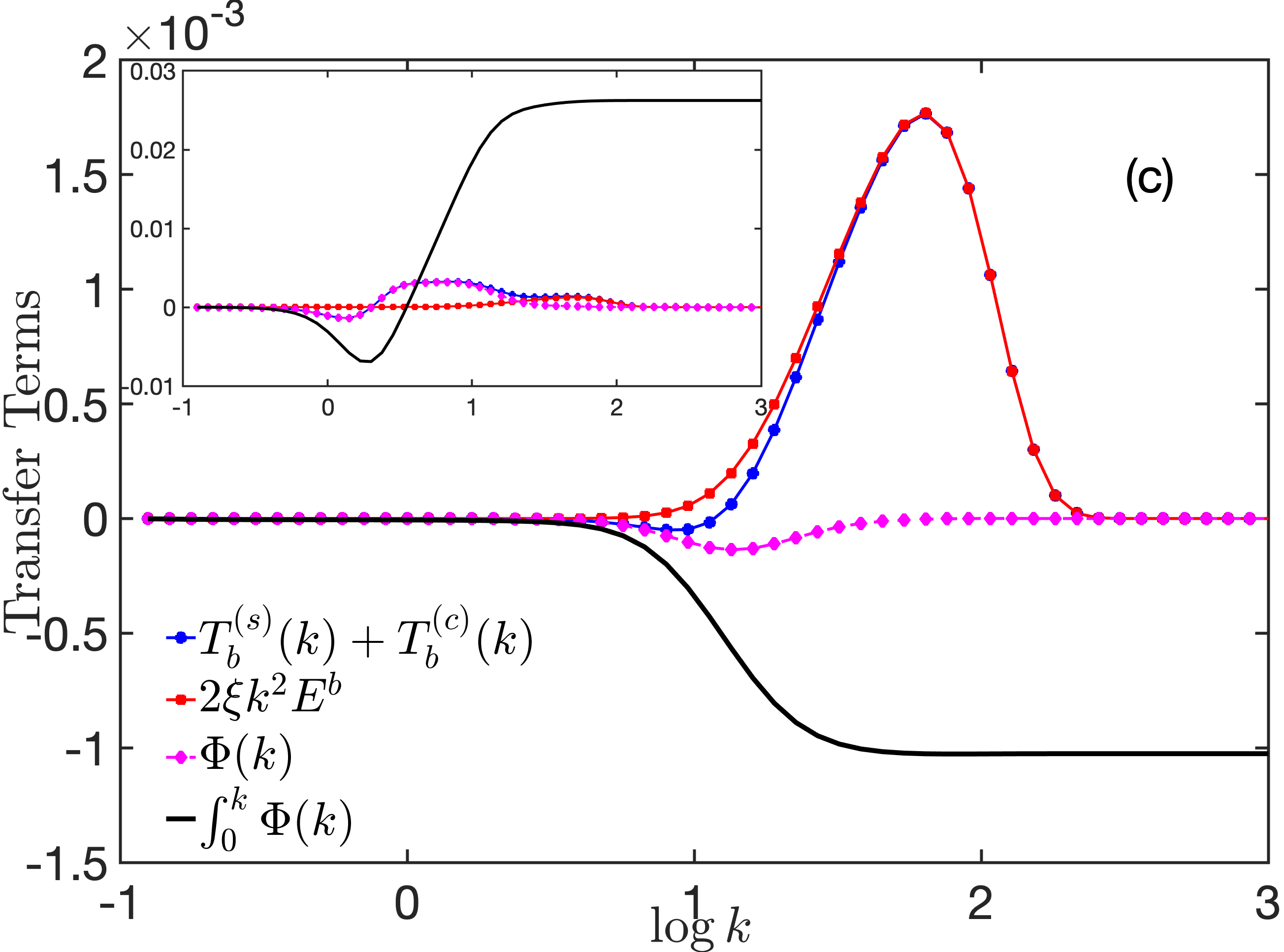}
\caption{Scale-by-scale plots of $T_b^{(s)}(k) +
  T_b^{(c)}(k)$, the effective magnetic diffusive term $2\eta k^2
	\ef{b}$, the difference $\Phi(k) \equiv T_b^{(s)}(k) + T_b^{(c)}(k) - 2\eta k^2
	\ef{b}$, and the cumulative sum of the difference $\int_0^k  \Phi(k)$  for  (a)
	$\dL \gtrsim d = 2.0$ (b) $\dL
  \lesssim d = 4.0 \lesssim \dU$ and (c) $\dU \lesssim d = 8.0$ at
	short $t = 1.0$ (inset) and long $t = 30.0$ times. The scale-by-scale behaviour
	of $\Phi(k)$ is a clear measure of the  effective  pumping  or  dissipation  of
	magnetic energy and the cumulative sum shows the net  effect  of  the  combined
	action of  the  transfer  terms  and  magnetic  dissipation.  The  insets  also
	underline why at short times there  is  always  an  increase  in  the  magnetic
	energy; at longer times, depending on the dimension, there is a net decrease or
	decrease of the same leading to a dimension-dependent dynamo---no-dynamo  phase
	diagram.}
\label{fig:transfer_B}
\end{figure*}

Is it possible to have a theoretical explanation, starting from the equations of
motion, which suggests  such  a  phase  diagram?  While  the  short  answer  is,
unfortunately, no, a scrutiny of  the  EDQNM-MHD  model  suggests  that  in  the
coupled set of equations, dynamo action for $\dL \leqslant d \leqslant \dU$  can
only be a consequence of a predominant energy transfer from ${\bf  u}  \to  {\bf
b}$, with the transfer term acting as  an  effective  forcing  on  the  magnetic
field. This preferential transfer of energy (at scales larger than  those  where
the diffusive damping becomes strong) leads to an increasing $R(t)$ followed  by
an eventual saturation stemming  from  the  nonlinearity  (negligible  at  short
times) and damping. Similarly, for $d < \dL$ or $d > \dU$ the large-scale energy
transfer ought to be, preferentially, from ${\bf b} \to {\bf u}$, even if  there
is a net ${\bf u} \to {\bf b}$ transfer at smaller scales. This  is  because  at
small scales the magnetic  dissipation  term  acts  as  a  counter  to  the  net
\emph{pumping} from the fluid field.

The argument outlined above is admittedly heuristic and a consequence of what we
see  in  Fig.~\ref{fig:dynamo_growth}.  The  only  way  to  make  this  argument
plausible is to numerically analyse the spectral properties of  the  interaction
terms in Eq.~\eqref{eq_main:transfer_split2}.  In  Fig.~\ref{fig:transfer_B}  we
plot,  scale-by-scale,  $\tn{b}{(s)}(k)  +  \tn{b}{(c)}(k)$  together  with  the
magnetic diffusion term $2\eta k^2 \ef{b}$ at (inset) short ($t = 1$)  and  long
($t = 30$) times for (a) $d  =  2.03$,  (b)  $d  =  4.0$  and  (c)  $d  =  8.0$.
Furthermore, we calculate and show the net transfer $\Phi(k) = \tn{b}{(s)}(k)  +
\tn{b}{(c)}(k) - 2\eta k^2 \ef{b}$ which is a clear  indication  of  the  scales
which leak ($\Phi(k) < 0$) or pile on ($\Phi(k) > 0$) magnetic energy.  However,
as Eq.~\eqref{eq:dynamo_rate_spectral} suggest, the dynamo action is essentially
an outcome of the integral of $\Phi(k)$; to make this point  succinct,  we  also
show in the same figure the cumulative integral $\int_0^k\Phi(k)$ as a  function
of the wavenumber $k$. Clearly, as $k \to \infty$, this  $\int_0^k\Phi(k)  >  0$
for $\dL \leqslant d \leqslant \dU$ and $\int_0^k\Phi(k) < 0$ for $d <  \dL$  or
$d > \dU$.

\begin{figure}[t!]\centering
\includegraphics[width=1.0\columnwidth]{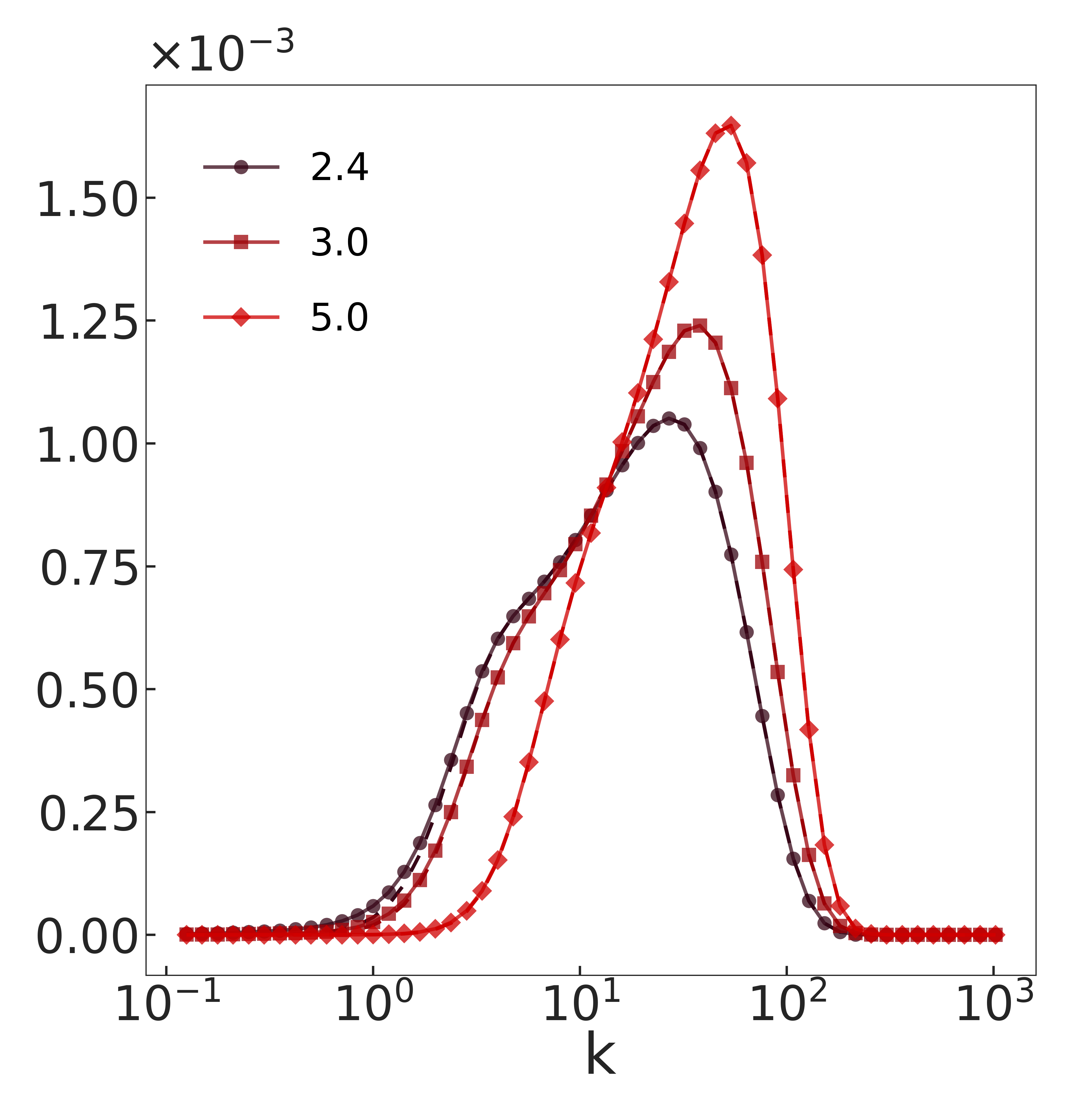}
\caption{The transfer $T_b^{(s)}(k) +
	T_b^{(c)}(k)$ (solid lines and symbols) and the magnetic dissipation $2\eta k^2
	\ef{b}$ (dashed lines) terms, for dimensions where dynamo action is  sustained,
	at a very long time $t = 100$. The nearly  indistinguishable  curves  for  each
	dimension is confirmation of the net balance between the ${\bf u} \to {\bf  b}$
	pumping and the magnetic dissipation leading to a saturation  of  the  magnetic
	energy and dynamo action.}
\label{fig:saturated}
\end{figure}

Figure~\ref{fig:transfer_B} is then a clear illustration of our  conjecture  and
consistent with observations in Fig.~\ref{fig:dynamo_growth}.  At  short  times,
the cumulative transfer (solid black line) is strictly \emph{forcing} leading to
an initial growth of the total magnetic energy.  At  long  times,  however,  the
situation is more delicate as the final state depends on the interaction of  the
fluid and magnetic components. Note that unlike kinematic models where the fluid
component (velocity field or kinetic energy spectrum) is frozen,  our  MHD-EQDNM
is able to go beyond the linear regime and provide a definitive  answer  to  the
dynamo problem. The final steady state of our MHD-EQDNM systems strongly depends
on the dimension. For $d = 4.0$ the net transfer is strictly positive leading to
dynamo action as seen in Fig.~\ref{fig:dynamo_growth}. A close inspection of the
net transfer $\Phi(k)$  and  its  cumulative  integral  underlines  this  effect
strongly. Furthermore,  the  small  scales  of  pumping  allow  for  a  lack  of
compensation from the diffusive term leading to growth of the  magnetic  energy.
Indeed,  for  such  dimensions  $\dL  \lesssim   d   \lesssim   \dU$,   we   see
(Fig.~\ref{fig:saturated})  that  at  long  times  there  is  a   scale-by-scale
cancellation of the pumping and damping leading to the  saturation  of  magnetic
energies and dynamo action.

%\begin{figure}[t!]\centering
%\includegraphics[width=1.0\columnwidth]{fig:transfer_B_saturated.png}
%\caption[Transfer \emph{vs} Diffusion terms for the dynamo effect]{The transfer terms (in
%symbols) and the magnetic diffusion term (in dashed) line
%for different dimensions at time $t = 100$. In this dynamo
%phase where the magnetic energy saturates to a finite value there is an exact
%compensation of the two contributions as discussed in the text.}
%\label{fig:transfer_B_saturated}
%\end{figure}

%\begin{figure}[t!]\centering
%\includegraphics[width=1.0\columnwidth]{fig:spectrum_B.png}
%\caption{Log-log plots of the
%  magnetic energy spectrum $\ef{b}$ vs $k$ for different dimensions.
%  The forcing scale from the fluid-magnetic interactions show up at
%  intermediate scales which become larger for $d \gtrsim \dU$. The scaling
%exponent exponent switches from -3 ($d \lesssim \dL$) to -5/3 ($d \gtrsim \dL$)
%as indicated by the dashed lines indicating power-laws as a guide to the eye.}
%\label{fig:spectrum_B}
%\end{figure}

However, for dimensions $d = 2.03$ and $d  =  8.0$  which  are  clearly  in  the
no-dynamo phase (Fig.~\ref{fig:dynamo_growth}), the spectral properties are more
involved. At low wavenumbers (with negligible  damping),  the  net  transfer  is
mainly from ${\bf b} \to {\bf u}$ leading to a  depletion  of  magnetic  energy.
While there is still a persistent net ${\bf u} \to {\bf b}$  transfer  for  such
dimensions, these happen at large wavenumbers (unlike what  is  seen  for  $d  =
4.0$) and hence damped out by the magnetic diffusivity. Such a spectral analysis
thus is useful in providing not a theory, but  an  understanding  of  where  the
dynamo--no-dynamo transition may happen as a function of the dimension  $d$.  In
particular, and as already suggested in Fig.~\ref{fig:dynamo_growth}, it clearly
shows the possibility of a lower $\dL$ and upper $\dU$ critical dimension,  tied
to the diffusive scales, marking out the boundary between dynamo and a no-dynamo
phase.

%Further evidence of this is implicit in the nature of the magnetic energy
%spectrum for different dimensions. In Fig.~\ref{fig:spectrum_B} we show
%log-log plots of the magnetic spectrum for several dimensions, both in the
%dynamo and the no-dynamo phase. We measure the spectra at time $t\gtrsim
%t^{\ast}$, where $t^{\ast}$ is the peak in $R(t)$. We find a transition in the
%spectral exponent on either side of the lower \emph{critical dimension}: For $d
%\lesssim \dL$, $\ef{b} \sim k^{-3}$ and for $d \gtrsim \dL$, $\ef{b} \sim
%k^{-5/3}$.  Furthermore, the peak in the spectrum corresponds to wavenumbers
%where the pumping via the transfer terms peak lending further evidence of the
%interpretation of the dynamo---no-dynamo phase transition outlined above.

\begin{figure}[t!]\centering
  \includegraphics[width=1.0\columnwidth]{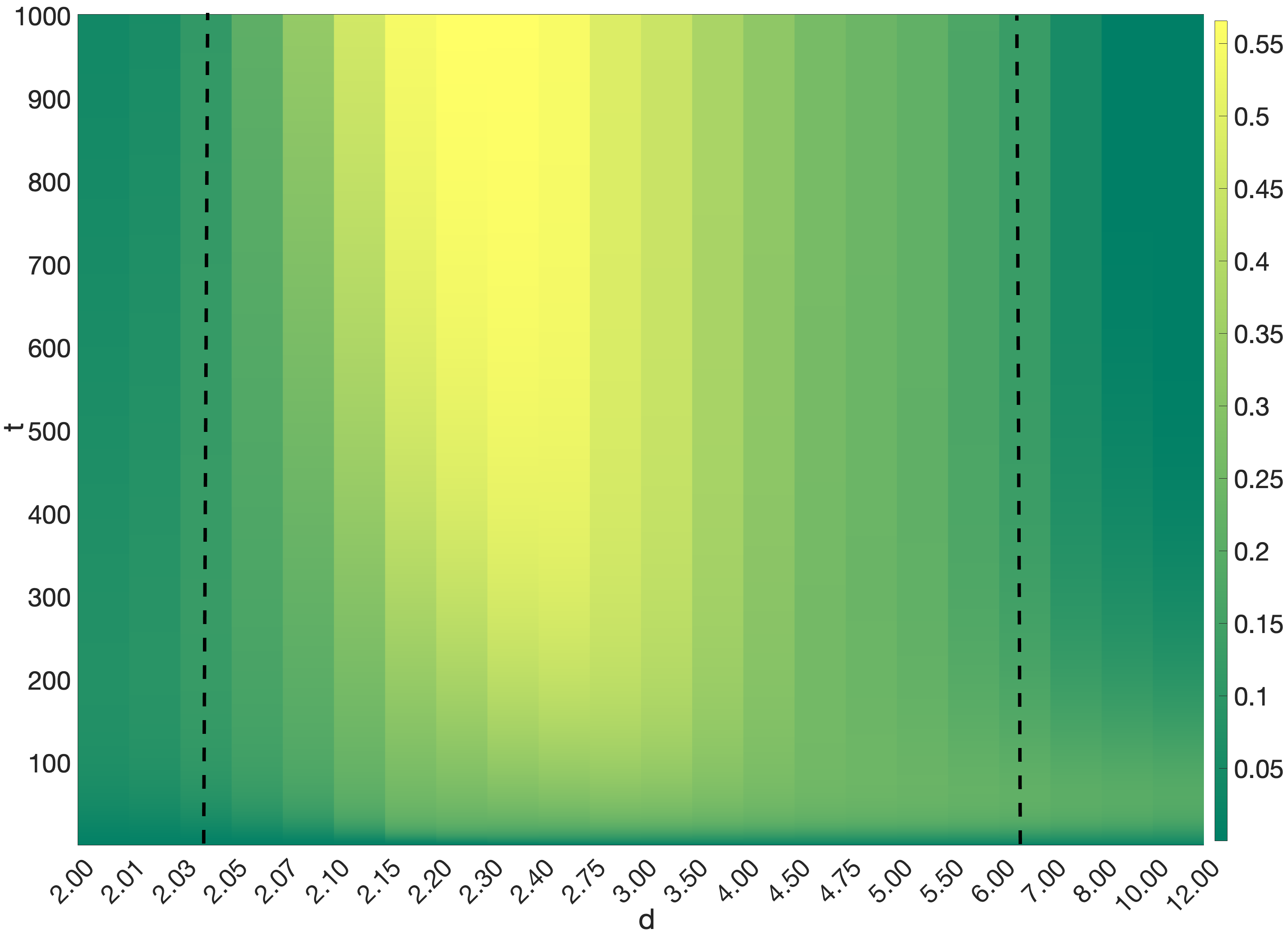}
  \caption[Pseudo-color  plot  of  Dynamo  effect  for   various   dimensions]{A
  \emph{space-time} color
    plot of the fraction $R(t)$ of magnetic to fluid energy. The dynamo phase
	(with colors ranging from light green to yellow) are
    indicated by thick,  black  vertical  dashed  lines  suggesting  lower  $\dL
    \approx 2.04$ and upper $\dU \approx  6.5$  \emph{critical}  dimensions  for
    dynamo action.}
\label{fig:dynamo_phaseplot}
\end{figure}

All  of  this  now  leads  us  to  construct   the   phase   diagram   for   the
dynamo---no-dynamo transition.  In  Fig.~\ref{fig:dynamo_phaseplot}  we  show  a
\textit{space-time} pseudo-color plot of $R(t)$ as a function of dimensions  and
time. Clearly at long times, $R(t) \to 0$ for $d \lesssim 2.04 \approx \dL$  and
for $d \gtrsim 6.5 \approx \dU$. These dimensions are indicated by the  vertical
broken lines in the figure and our numerical simulations of the MHD-EDQNM  model
predicts dynamo action for all dimensions which lie in between these two.

In this paper, we have focussed on showing the existence of a dynamo---no-dynamo
phase boundary, for a given point in the magnetic  Reynolds  number  and  Prandl
number  landscape,   by   constructing   a   $d$-dimensional   MHD-EDQNM   model
(Eqs.~\eqref{eq_main:edqnm-mhd_spectrum}~--~\eqref{eq_main:transfer_split2}). It
is important to stress that  in  the  absence  of  a  theoretical  estimate  the
\textit{precise} value of $\dL$ and $\dU$ is moot; however,  the  existence  of
such a lower dimension greater than $d = 2.0$ and, more surprisingly,  an  upper
dimension makes this study intriguing. Furthermore,  our  $d$-dimensional  model
can be used to investigate a wide range of Prandtl and magnetic Reynolds numbers
which are currently difficult  in  full  MHD  direct  numerical  simulation.  In
particular, in the kinematic regime, the form of the energy
spectrum can be prescribed or modified in such a way as  to  include  a
non-zero    helicity     and     cross-helicity     ~\cite{briard_dynamics_2017,
briard_decay_2018,   turner_eddy-damped_2002},   effects   of   compressibility
\cite{martins_afonso_kazantsev_2019} or even to vary the spatial regularity
of the velocity field \cite{vincenzi_kraichnankazantsev_2002}, which  could,  in
principle, be addressed for any given $d$. We hope that
this model will trigger further interest in tackling the important questions  of
dynamo from a firmer theoretical standpoint with a greater emphasis on the  role
of triadic interactions.

%HHHHHHHHHHHHHHHHHHHHHHHHHHHHHHHHHHHHHHHHHHHHHHHHHHHHHHHHHHHHHHHHHHHHHHHHHHHHHH
\begin{acknowledgements}

	S.D.M, S.S.R. and D.V. thank the Indo–French Centre for Applied
	Mathematics (IFCAM) for financial support.  D.V. acknowledges his
	Associateship with the International Centre for Theoretical Sciences,
	Tata Institute of Fundamental Research, Bangalore, India.  The
	simulations were performed on the ICTS clusters \emph{Tetris} and
	\emph{Contra}. SSR acknowledges SERB-DST (India) projects
	STR/2021/000023 and CRG/2021/002766 for financial support and would
	like to thank the Isaac Newton Institute for Mathematical Sciences,
	Cambridge, for support and hospitality during the programme
	\textit{Anti-diffusive dynamics: from sub-cellular to astrophysical
	scales} (EPSRC grant EP/R014604/1), where part of the work on this
	paper was undertaken. This research was supported in part by the
	International Centre for Theoretical Sciences (ICTS) for participating
	in the programs ---  \textit{Field Theory and Turbulence}
	(code:ICTS/ftt2023/12) and \textit{Turbulence: Problems at the
	Interface of Mathematics and Physics} (code: ICTS/TPIMP2020/12).  SSR
	acknowledges the support of the DAE, Govt. of India, under project no.
	12-R\&D-TFR-5.10-1100 and project no. RTI4001.

\end{acknowledgements}
%HHHHHHHHHHHHHHHHHHHHHHHHHHHHHHHHHHHHHHHHHHHHHHHHHHHHHHHHHHHHHHHHHHHHHHHHHHHHHH
% \bibliographystyle{naturemag}
 \bibliographystyle{unsrt}
 \bibliography{References}

\begin{thebibliography}{10}

\bibitem{ogilvie_astrophysical_2016}
Gordon~I. Ogilvie.
\newblock Astrophysical fluid dynamics.
\newblock {\em Journal of Plasma Physics}, 82(3):205820301, June 2016.

\bibitem{brandenburg_advances_2018}
Axel Brandenburg.
\newblock Advances in mean-field dynamo theory and applications to astrophysical turbulence.
\newblock {\em Journal of Plasma Physics}, 84(4):735840404, 2018.

\bibitem{brandenburg_galactic_2023}
Axel Brandenburg and Evangelia Ntormousi.
\newblock Galactic dynamos.
\newblock {\em Annual Review of Astronomy and Astrophysics}, 61:561--606, 2023.

\bibitem{moffatt_selfexciting_2019}
Keith Moffatt and Emmanuel Dormy.
\newblock {\em Self-Exciting Fluid Dynamos}.
\newblock Cambridge University Press, Cambridge, UK, 2019.

\bibitem{rincon_dynamo_2019}
Fran{\c{c}}ois Rincon.
\newblock Dynamo theories.
\newblock {\em Journal of Plasma Physics}, 85(4):205850401, August 2019.

\bibitem{tobias_turbulent_2021}
S.~M. Tobias.
\newblock The turbulent dynamo.
\newblock {\em Journal of Fluid Mechanics}, 912:P1, April 2021.

\bibitem{choudhuri_physics_1998}
Arnab~Rai Choudhuri.
\newblock {\em The {Physics} of {Fluids} and {Plasmas}: {An} {Introduction} for {Astrophysicists}}.
\newblock Cambridge University Press, 1998.

\bibitem{davidson_introduction_2001}
P.~A. Davidson.
\newblock {\em An {Introduction} to {Magnetohydrodynamics}}.
\newblock Cambridge University Press, 2001.

\bibitem{biskamp_magnetohydrodynamic_2003}
Dieter Biskamp.
\newblock {\em Magnetohydrodynamic {Turbulence}}.
\newblock Cambridge University Press, 2003.

\bibitem{galtier_introduction_2016}
Sebastien Galtier.
\newblock {\em Introduction to Modern Magnetohydrodynamics}.
\newblock Cambridge University Press, Cambridge, UK, 2016.

\bibitem{schekochihin_mhd_2022}
Alexander~A. Schekochihin.
\newblock {MHD} turbulence: a biased review.
\newblock {\em Journal of Plasma Physics}, 88:155880501, 2022.

\bibitem{monchaux_generation_2007}
R.~Monchaux, M.~Berhanu, M.~Bourgoin, M.~Moulin, Ph. Odier, J.-F. Pinton, R.~Volk, S.~Fauve, N.~Mordant, F.~P\'etr\'elis, A.~Chiffaudel, F.~Daviaud, B.~Dubrulle, C.~Gasquet, L.~Mari\'e, and F.~Ravelet.
\newblock Generation of a magnetic field by dynamo action in a turbulent flow of liquid sodium.
\newblock {\em Physical Review Letters}, 98:044502, Jan 2007.

\bibitem{kazantsev_enhancement_1968}
A.~P. Kazantsev.
\newblock Enhancement of a {Magnetic} {Field} by a {Conducting} {Fluid}.
\newblock {\em Soviet Journal of Experimental and Theoretical Physics}, 26:1031, May 1968.

\bibitem{falkovich_particles_2001}
G.~Falkovich, K.~Gaw{\c{e}}dzki, and M.~Vergassola.
\newblock Particles and fields in fluid turbulence.
\newblock {\em Reviews of Modern Physics}, 73(4):913--975, November 2001.

\bibitem{brandenburg_astrophysical_2005}
Axel Brandenburg and Kandaswamy Subramanian.
\newblock Astrophysical magnetic fields and nonlinear dynamo theory.
\newblock {\em Physics Reports}, 417(1):1--209, October 2005.

\bibitem{gruzinov_small-scale-field_1996}
A.~Gruzinov, S.~Cowley, and R.~Sudan.
\newblock Small-{Scale}-{Field} {Dynamo}.
\newblock {\em Physical Review Letters}, 77(21):4342--4345, November 1996.

\bibitem{schekochihin_spectra_2002}
Alexander~A. Schekochihin, Stanislav~A. Boldyrev, and Russell~M. Kulsrud.
\newblock Spectra and {Growth} {Rates} of {Fluctuating} {Magnetic} {Fields} in the {Kinematic} {Dynamo} {Theory} with {Large} {Magnetic} {Prandtl} {Numbers}.
\newblock {\em The Astrophysical Journal}, 567(2):828, March 2002.

\bibitem{arponen_dynamo_2007}
Heikki Arponen and Peter Horvai.
\newblock Dynamo {Effect} in the {Kraichnan} {Magnetohydrodynamic} {Turbulence}.
\newblock {\em Journal of Statistical Physics}, 129(2):205--239, October 2007.

\bibitem{martins_afonso_kazantsev_2019}
Marco Martins~Afonso, Dhrubaditya Mitra, and Dario Vincenzi.
\newblock Kazantsev dynamo in turbulent compressible flows.
\newblock {\em Proceedings of the Royal Society A: Mathematical, Physical and Engineering Sciences}, 475(2223):20180591, March 2019.

\bibitem{wilson_critical_1972}
Kenneth~G. Wilson and Michael~E. Fisher.
\newblock Critical exponents in 3.99 dimensions.
\newblock {\em Physical Review Letters}, 28:240--243, Jan 1972.

\bibitem{fournier_d-dimensional_1978}
Jean-Daniel Fournier and Uriel Frisch.
\newblock d-dimensional turbulence.
\newblock {\em Physical Review A}, 17(2):747--762, February 1978.

\bibitem{lvov_quasi-gaussian_2002}
Victor~S. L'vov, Anna Pomyalov, and Itamar Procaccia.
\newblock Quasi-gaussian statistics of hydrodynamic turbulence in $\frac{4}{3}+ \epsilon$ dimensions.
\newblock {\em Physical Review Letters}, 89(6):064501, jul 2002.

\bibitem{celani_turbulence_2010}
Antonio Celani, Stefano Musacchio, and Dario Vincenzi.
\newblock Turbulence in {More} than {Two} and {Less} than {Three} {Dimensions}.
\newblock {\em Physical Review Letters}, 104(18):184506, May 2010.

\bibitem{frisch_turbulence_2012}
Uriel Frisch, Anna Pomyalov, Itamar Procaccia, and Samriddhi~Sankar Ray.
\newblock Turbulence in {Noninteger} {Dimensions} by {Fractal} {Fourier} {Decimation}.
\newblock {\em Physical Review Letters}, 108(7):074501, February 2012.

\bibitem{ray_thermalized_2015}
Samriddhi~Sankar Ray.
\newblock Thermalized solutions, statistical mechanics and turbulence: {An} overview of some recent results.
\newblock {\em Pramana}, 84(3):395--407, mar 2015.

\bibitem{ray_non-intermittent_2018}
Samriddhi~Sankar Ray.
\newblock Non-intermittent turbulence: {Lagrangian} chaos and irreversibility.
\newblock {\em Physical Review Fluids}, 3(7):072601, July 2018.

\bibitem{picardo_lagrangian_2020}
Jason~R. Picardo, Akshay Bhatnagar, and Samriddhi~Sankar Ray.
\newblock Lagrangian irreversibility and {Eulerian} dissipation in fully developed turbulence.
\newblock {\em Physical Review Fluids}, 5(4):042601, April 2020.

\bibitem{kraichnan_inertial_1967}
Robert~H. Kraichnan.
\newblock Inertial {Ranges} in {Two-Dimensional} {Turbulence}.
\newblock {\em The Physics of Fluids}, 10(7):1417--1423, July 1967.

\bibitem{orszag_analytical_1970}
Steven~A. Orszag.
\newblock Analytical theories of turbulence.
\newblock {\em Journal of Fluid Mechanics}, 41(2):363--386, April 1970.

\bibitem{rose_ha_fully_1978}
{Rose, H.A.} and {Sulem, P.L.}
\newblock Fully developed turbulence and statistical mechanics.
\newblock {\em Journal de Physique}, 39(5):441--484, 1978.

\bibitem{clark_effect_2021}
Daniel Clark, Richard D. J.~G. Ho, and Arjun Berera.
\newblock Effect of spatial dimension on a model of fluid turbulence.
\newblock {\em Journal of Fluid Mechanics}, 912:A40, April 2021.

\bibitem{clark_critical_2022}
Daniel Clark, Andres Armua, Richard D. J.~G. Ho, and Arjun Berera.
\newblock Critical transition to a non-chaotic regime in isotropic turbulence.
\newblock {\em Journal of Fluid Mechanics}, 930:A17, 2022.

\bibitem{lesieur_turbulence_2008}
Marcel Lesieur.
\newblock {\em Turbulence in Fluids}.
\newblock Springer Netherlands, 2008.

\bibitem{pouquet_strong_1976}
A.~Pouquet, U.~Frisch, and J.~Leorat.
\newblock Strong {MHD} helical turbulence and the nonlinear dynamo effect.
\newblock {\em Journal of Fluid Mechanics}, 77:321--354, September 1976.

\bibitem{schilling_triadic_2002}
Oleg Schilling and Ye~Zhou.
\newblock Triadic energy transfers in non-helical magnetohydrodynamic turbulence.
\newblock {\em Journal of Plasma Physics}, 68(5):389--406, November 2002.

\bibitem{pouquet_two-dimensional_1978}
Annick Pouquet.
\newblock On two-dimensional magnetohydrodynamic turbulence.
\newblock {\em Journal of Fluid Mechanics}, 88:1--16, September 1978.

\bibitem{frisch_crossover_1976}
U.~Frisch, M.~Lesieur, and P.~L. Sulem.
\newblock Crossover {Dimensions} for {Fully} {Developed} {Turbulence}.
\newblock {\em Physical Review Letters}, 37(14):895--897, October 1976.

\bibitem{zeldovich_magnetic_1957}
Ia.~B. Zel'dovich.
\newblock The magnetic field in the two-dimensional motion of a conducting turbulent liquid.
\newblock {\em Sov. Phys. JETP}, 4:460--462, 1957.

\bibitem{zeldovich_magnetic_1980_jetp}
Ya.~B. Zel'dovich and A.~A. Ruzma\u{\i}kin.
\newblock The magnetic field in a conducting fluid in two-dimensional motion.
\newblock {\em Sov. Phys. JETP}, 51:493--497, 1980.

\bibitem{briard_dynamics_2017}
Antoine Briard and Thomas Gomez.
\newblock Dynamics of helicity in homogeneous skew-isotropic turbulence.
\newblock {\em Journal of Fluid Mechanics}, 821:539--581, June 2017.

\bibitem{briard_decay_2018}
Antoine Briard and Thomas Gomez.
\newblock The decay of isotropic magnetohydrodynamics turbulence and the effects of cross-helicity.
\newblock {\em Journal of Plasma Physics}, 84(1):905840110, February 2018.

\bibitem{turner_eddy-damped_2002}
Leaf Turner and Jane Pratt.
\newblock Eddy-damped quasinormal {Markovian} closure: a closure for magnetohydrodynamic turbulence?
\newblock {\em Journal of Physics A: Mathematical and General}, 35(3):781--793, January 2002.

\bibitem{vincenzi_kraichnankazantsev_2002}
D.~Vincenzi.
\newblock The {Kraichnan}–{Kazantsev} {Dynamo}.
\newblock {\em Journal of Statistical Physics}, 106(5):1073--1091, March 2002.

\bibitem{sagaut_edqnm_2002}
Pierre Sagaut.
\newblock {EDQNM} {Modeling}.
\newblock In Pierre Sagaut, editor, {\em Large {Eddy} {Simulation} for {Incompressible} {Flows}: {An} {Introduction}}, pages 391--395. Springer, 2002.

\bibitem{sagaut_homogeneous_2018}
Pierre Sagaut and Claude Cambon.
\newblock {\em Homogeneous {Turbulence} {Dynamics}}.
\newblock Springer International Publishing, 2018.

\end{thebibliography}

\onecolumngrid

\section{Appendix} 
\renewcommand{\theequation}{\arabic{equation}}
\setcounter{equation}{0}  % reset counter 

The governing \mhd equations for the unit density incompressible velocity $\vb{u}$ ($\nabla \cdot \vb{u} = 0$) and magnetic $\vb{b}$
($\nabla \cdot \vb{b} = 0$) fields are
\begin{subequations}
\begin{align}
  \partial_{t}\vb{u} &= -\grad{P} - \qty(\vb{u}\cdot \bm{\nabla })\vb{u} + \qty(\vb{b}\cdot \bm{\nabla }) \vb{b} + \nu \nabla ^2 \vb{u}; \label{eq:V_evolution_real} \\
  \partial_{t}\vb{b} &= \curl{\qty(\vb{u}\cp\vb{b}) } + \eta  \nabla ^2 \vb{b}. \label{eq:B_evolution_real}
\end{align}
\label{eq:evolution_real}%
\end{subequations}
The pressure field is given by $P$, the kinematic fluid viscosity is $\nu$, and
the magnetic diffusivity is $\eta$.  The kinetic helicity $\vb{u}\vdot\bm{\omega
}$, magnetic helicity $\vb{A}\vdot\vb{b}$, with the magnetic potential $\vb{A}$
defined via $\vb{b}=\curl{\vb{A}}$, and cross helicity $\vb{u}\vdot\vb{b}$ are
all assumed to be zero for all times.

In Appendices A-C we give a detailed derivation of the $d$-dimensional MHD-EDQNM equations going through 
the successive approximations. A complete numerical prescription to solve these equations is found in 
Appendix D.

\section{Appendix A: The Quasi-Normal Approximation} 

\renewcommand{\theequation}{A-\arabic{equation}}
\setcounter{equation}{0}  % reset counter 

The derivation of the closure model follows best from the Fourier space
representation of the MHD equations, expressed conveniently in a symmetric form
between the fluid and magnetic fields, written in component form with Greek
indices: 
\begin{subequations}
\begin{align}
  \qty[\dv{}{t}+\nu k^2] u_{\alpha }\qty(\vb{k},t)&= \Pc{k}_{\alpha \rho \gamma } \Iq \Ip \qty[ u_{\rho }(\vb{q})u_{\gamma }(\vb{p})-b_{\rho }(\vb{q})b_{\gamma }(\vb{p})] \bm{\delta }\qty(\vb{p}+\vb{q}-\vb{k}); \label{eq:V_evolution_spectral}\\
  \qty[\dv{}{t}+\eta k^2] b_{\alpha }\qty(\vb{k},t)&= \Cc{k}_{\alpha \rho \gamma } \Iq \Ip \qty[ b_{\rho }(\vb{q})u_{\gamma }(\vb{p})+ b_{\rho }(\vb{p})u_{\gamma }(\vb{q})]\bm{\delta }\qty(\vb{p}+\vb{q}-\vb{k}). \label{eq:B_evolution_spectral}
\end{align}
\label{eq:evolution_spectral}
\end{subequations}
By defining $\Pb{k}_{\alpha \beta }= \delta_{\alpha \beta }- \dfrac{k_{\alpha }k_{\beta }}{k^2}$, we obtain 
\begin{subequations}
\begin{align}
\Pc{k}_{\alpha \rho \gamma }&=-\frac{\iota }{2}\qty(\Pb{k}_{\alpha \rho }k_{\gamma }+ \Pb{k}_{\alpha \gamma }k_{\rho }); \label{eq:projection_tensor} \\
\Cc{k}_{\alpha \rho \gamma }&=-\frac{\iota }{2}\qty(\delta _{\alpha \rho }k_{\gamma }-\delta_{\alpha \gamma }k_{\rho }) \label{eq:transport_tensor}
\end{align}
\end{subequations}
for the project and transport tensors, respectively.

The form of the generalised $n^{\mathrm{th}}$ order spectral moment (for fields $X^{i}$) 
\begin{align}
  \zf{X^{(1)}X^{(2)}\cdots X^{(n)}}_{\alpha _1\alpha _2 \cdots \alpha _{n}} \qty(k_1,k_2, \dots, k_{n-1} )
  &=  \ev{X^{(1)}_{\alpha _1}\qty(k_1) X^{(2)}_{\alpha _2}\qty(k_2) \cdots X^{(n-1)}_{\alpha _{n-1}}\qty(k_{n-1}) X^{(n)}_{\alpha _n}\qty(- \sum_{i=1}^{n-1} \; k_i) }
\label{eq:n-order_mom}
\end{align}
allows us to obtain the evolution equations
for the second moments $\zf{uu}$ and $\zf{bb}$:
\begin{subequations}
\begin{align}
  \dop{2\nu k^2}\zf{uu}_{\alpha \beta }(k) &= \Itr \Pc{-k}_{\beta  \rho \gamma }\qty[\zf{uuu}_{\alpha  \rho \gamma }(k,-q)-\zf{ubb}_{\alpha  \rho \gamma }(k,-q)] + \eval{\mathrm{c}\cdot \mathrm{c}}_{\alpha \leftrightarrow \beta  }; \label{eq:2-order_mom_V2} \\
  \dop{2\eta k^2}\zf{bb}_{\alpha \beta }(k)&=  \Itr \Cc{-k}_{\beta  \rho \gamma }\qty[\zf{bbu}_{\alpha  \rho \gamma }(k,-q)+\zf{bbu}_{\alpha \rho \gamma }(k,-p)] + \eval{\mathrm{c}\cdot\mathrm{c}}_{\alpha \leftrightarrow \beta }. \label{eq:2-order_mom_B2}
\end{align}
\label{eq:2-order_mom}
\end{subequations}
Here  $\Itr \equiv \Iq \Ip \delta \qty(\vb{q}+\vb{p}-\vb{k})$,
$\mathrm{c}\cdot \mathrm{c}$ denotes complex conjugate, and $\alpha
\leftrightarrow \beta$ implies the exchange of indices.

Similarly, the evolution of the third-order moments $\zf{uuu},\zf{ubb}$, and $\zf{bbu}$ follows:
\begin{subequations}
\begin{align}
  \dop{\nu \qty(k^2+q^2+p^2)} \zf{uuu}_{\alpha  \rho \gamma }(k,-q)&= \Ir \Pc{k}_{\alpha \mu \sigma }\qty[\zf{uuuu}_{\rho \gamma \mu \sigma }(-q,-p,r)-\zf{uubb}_{\rho \gamma \mu \sigma }(-q,-p,r)] \nonumber \\
    &\;+ \Ir \Pc{-q}_{\rho \mu \sigma }\qty[ \zf{uuuu}_{\alpha \gamma \mu \sigma }(k,-p,r)-\zf{uubb}_{\alpha \gamma \mu \sigma }(k,-p,r)]\nonumber \\
  &\; + \Ir \Pc{-p}_{\gamma \mu \sigma }\qty[ \zf{uuuu}_{\alpha \rho \mu \sigma }(k,-q,r)- \zf{uubb}_{\alpha \rho \mu \sigma }(k,-q,r)]
  \label{eq:3-order_mom_V}\\
  \dop{\nu k^2+\eta \qty(q^2+ p^2)} \zf{ubb}_{\alpha \rho \gamma }(k,-q)&= \Ir \Pc{k}_{\alpha \mu \sigma }\qty[\zf{bbuu}_{\rho \gamma \mu \sigma }(-q,-p,r)-\zf{bbbb}_{\rho \gamma \mu \sigma }(-q,-p,r)] \nonumber \nonumber \\
&\; + 2\Ir \qty[\Cc{-q}_{\rho \mu \sigma }\zf{bbuu}_{\gamma \mu \alpha \sigma }(-p,r,k)+ \Cc{-p}_{\gamma \mu \sigma }\zf{bbuu}_{\rho \mu \alpha \sigma }(-q,r,k) ]
  \label{eq:3-order_mom_B}\\
  \dop{\nu p^2+\eta \qty(k^2+q^2)}\zf{bbu}_{\alpha \rho \gamma }(k,-q)&= \Ir \Pc{-p}_{\gamma \mu \sigma }\qty[\zf{bbuu}_{\alpha \rho \mu \sigma }(k,-q,r)-\zf{bbbb}_{\alpha \rho \mu \sigma }(k,-q,r)]\nonumber \\
  &\; + 2\Ir \qty[\Cc{k}_{\alpha \mu \sigma }\zf{bbuu}_{\rho \mu \gamma \sigma }(-q,r,-p)+\Cc{-q}_{\rho \mu \sigma }\zf{bbuu}_{\alpha \mu \gamma \sigma }(k,r,-p)] \label{eq:3-order_mom_B2}\\
  \dop{\nu q^2+\eta \qty(k^2+p^2)}\zf{bbu}_{\alpha \rho \gamma }(k,-p)&= \Ir \Pc{-q}_{\gamma \mu \sigma }\qty[\zf{bbuu}_{\alpha \rho \mu \sigma }(k,-p,r)-\zf{bbbb}_{\alpha \rho \mu \sigma }(k,-p,r)]\nonumber \\
  &\; + 2\Ir \qty[\Cc{k}_{\alpha \mu \sigma }\zf{bbuu}_{\rho \mu \gamma \sigma }(-p,r,-q)+\Cc{-p}_{\rho \mu \sigma }\zf{bbuu}_{\alpha \mu \gamma \sigma }(k,r,-q)].
  \label{eq:3-order_mom_B3}
\end{align}%
\label{eq:3-order_mom}%
\end{subequations}%
A comparison between Eqs.~\eqref{eq:2-order_mom} and \eqref{eq:3-order_mom}
underlines the closure problem inherent in such models: Solving for the
$n^{\mathrm{th}}$ moment is contingent on knowing the $(n+1)^{\mathrm{th}}$ moment.
Hence suitable approximations are needed to close this hierarchy and find, for our problem, 
a closed form representation of the second-order moments. One such approach 
is the Quasi-Normal approximation which assumes that the statistics to be essentially Gaussian 
(with a vanishing cumulant) and hence
  \begin{align}
    \zf{X^{(1)}X^{(2)}X^{(3)}X^{(4)}}_{\alpha \beta \rho \gamma }(k_1,k_2,k_3)&=
    \zf{X^{(1)}X^{(2)}}_{\alpha \beta }(k_1) \zf{X^{(3)}X^{(4)}}_{\rho \gamma  }(k_3) \delta (k_1+k_2)+
    \zf{X^{(1)}X^{(3)}}_{\alpha \rho  }(k_1) \zf{X^{(2)}X^{(4)}}_{\beta \gamma }(k_2) \delta (k_1+k_3)\nonumber \\
    &+\zf{X^{(1)}X^{(4)}}_{\alpha \gamma}(k_1) \zf{X^{(2)}X^{(3)}}_{\beta \rho   }(k_2)\delta (k_2+k_3).
    \label{eq:quasi-normal}%
  \end{align}%
This form allows us (with the further assumption $\ev{u_{\alpha }(k_1) b_{\beta }(k_2)}=0$) to reduce Eq.~\eqref{eq:3-order_mom} 
to 
\begin{subequations}
  \begin{align}
  \dop{\nu \qty(k^2+p^2+q^2)}\zf{uuu}_{\alpha \rho \gamma }\qty(k,-q)&=  2\left[\Pc{k}_{\alpha \mu \sigma }\zf{uu}_{\rho \mu }(-q)\zf{uu}_{\gamma \sigma }(-p)+\Pc{-q}_{\rho \mu \sigma } \zf{uu}_{\alpha \sigma }(k)\zf{uu}_{\gamma \mu }(-p) \right. \nonumber \\
  & + \left. \Pc{-p}_{\gamma \mu \sigma }\zf{uu}_{\alpha \mu }(k)\zf{uu}_{\rho \sigma }(-q)\right] \label{eq:3-order_mom_V_qn}\\
\dop{\nu k^2+\eta \qty(q^2+ p^2)} \zf{ubb}_{\alpha \rho \gamma }(k,-q)&=2 \left[ \Cc{-q}_{\rho \mu \sigma }\zf{bb}_{\gamma \mu }(-p)\zf{uu}_{\alpha \sigma }(k)+\Cc{-p}_{\gamma \mu \sigma }\zf{bb}_{\rho \mu }(-q)\zf{uu}_{\alpha \sigma }(k) \right. \nonumber \\
& - \left. \Pc{k}_{\alpha \mu \sigma }\zf{bb}_{\rho \mu }(-q)\zf{bb}_{\gamma \sigma }(-p) \right]
    \label{eq:3-order_mom_B_qn}\\
  \dop{\nu p^2+\eta \qty(k^2+q^2)} \zf{bbu}_{\alpha \rho \gamma }(k,-q)&=2\left[\Cc{k}_{\alpha \mu \sigma }\zf{bb}_{\rho \mu }(-q)\zf{uu}_{\gamma \sigma }(-p)+\Cc{-q}_{\rho \mu \sigma }\zf{bb}_{\alpha \mu }(k)\zf{uu}_{\gamma \sigma }(-p) \right. \nonumber \\
  & \left.- \Pc{-p}_{\gamma \mu \sigma }\zf{bb}_{\alpha \mu }(k)\zf{bb}_{\rho \sigma }(-q)\right]
    \label{eq:3-order_mom_B2_qn}\\
  \dop{\nu q^2+\eta \qty(k^2+p^2)} \zf{bbu}_{\alpha \rho \gamma }(k,-p)&=2\left[ \Cc{k}_{\alpha \mu \sigma }\zf{bb}_{\rho \mu }(-p)\zf{uu}_{\gamma \sigma }(-q)+\Cc{-p}_{\rho \mu \sigma }\zf{bb}_{\alpha \mu }(k)\zf{uu}_{\gamma \sigma }(-q) \right. \nonumber \\
  & -\left. \Pc{-q}_{\gamma \mu \sigma }\zf{bb}_{\alpha \mu }(k)\zf{bb}_{\rho \sigma }(-p) \right].
    \label{eq:3-order_mom_B3_qn}
  \end{align}
  \label{eq:3-order_mom_qn}%
\end{subequations}
This form allows us, by defining 
\begin{equation}
\dop{\omega }^{-1}\cdot=\bm{\hat{\theta }}(\omega )\, \cdot = \displaystyle \int_{0}^{t}\dd{s}\cdot e^{-(t-s)\omega },
\label{eq:def_inverse}
\end{equation}
to invert Eq.~\eqref{eq:3-order_mom_qn} and, on substitution in Eq.~\eqref{eq:2-order_mom}, obtain 
\begin{subequations}
 \begin{align}
    \dop{2\nu k^2}\zf{uu}_{\alpha \beta }(k)&= \Itr 2\Oth{kpq}\Pc{-k}_{\beta \rho \gamma }
    \left[\Pc{k}_{\alpha \mu \sigma }\zf{uu}_{\rho \mu }(-q)\zf{uu}_{\gamma \sigma }(-p)+
    \Pc{-q}_{\rho \mu \sigma } \zf{uu}_{\alpha \sigma }(k)\zf{uu}_{\gamma \mu }(-p) \right. \nonumber \\
    & \hspace{8.5em} \left. \; +\Pc{-p}_{\gamma \mu \sigma }\zf{uu}_{\alpha \mu }(k)\zf{uu}_{\rho \sigma }(-q) \right] + \eval{\mathrm{c}\cdot\mathrm{c}}_{\alpha \leftrightarrow \beta } \nonumber \\
   & - \Itr 2\OthB{kpq}\Pc{-k}_{\beta \rho \gamma } \left[ \Cc{-q}_{\rho \mu \sigma }\zf{uu}_{\alpha \sigma }(k)\zf{bb}_{\gamma \mu }(-p)+\Cc{-p}_{\gamma \mu \sigma }\zf{uu}_{\alpha \sigma }(k)\zf{bb}_{\rho \mu }(-q)  \right. \nonumber \\
   & \hspace{6.5em} \left. \; -\Pc{k}_{\alpha \mu \sigma }\zf{bb}_{\rho \mu }(-q)\zf{bb}_{\gamma \sigma }(-p) \right] + \eval{\mathrm{c}\cdot\mathrm{c}}_{\alpha \leftrightarrow \beta }
    \label{eq:2-order_mom_V3}\\
    \dop{2\eta k^2}\zf{bb}_{\alpha \beta }(k)&= \Itr  2\OthB{pkq} \Cc{-k}_{\beta \rho \gamma }\left[ \Cc{k}_{\alpha \mu \sigma }\zf{bb}_{\rho \mu }(-q)\zf{uu}_{\gamma \sigma }(-p)+\Cc{-q}_{\rho \mu \sigma }\zf{bb}_{\alpha \mu }(k)\zf{uu}_{\gamma \sigma }(-p)\right. \nonumber \\
   & \hspace{6.5em} \left. - \Pc{-p}_{\gamma \mu \sigma }\zf{bb}_{\alpha \mu }(k)\zf{bb}_{\rho \sigma }(-q) \right] \nonumber \\
   & \; + \Itr  2\OthB{qkp} \Cc{-k}_{\beta \rho \gamma }\left[ \Cc{k}_{\alpha \mu \sigma }\zf{bb}_{\rho \mu }(-p)\zf{uu}_{\gamma \sigma }(-q)+\Cc{-p}_{\rho \mu \sigma }\zf{bb}_{\alpha \mu }(k)\zf{uu}_{\gamma \sigma }(-q)\right. \nonumber \\
   & \hspace{6.5em} \left. - \Pc{-q}_{\gamma \mu \sigma }\zf{bb}_{\alpha \mu }(k)\zf{bb}_{\rho \sigma }(-p)
   \right] + \eval{\mathrm{c}\cdot\mathrm{c}}_{\alpha \leftrightarrow \beta }.
    \label{eq:2-order_mom_B3}
  \end{align}
\label{eq:2-order_mom_qn}%
\end{subequations}
The frequencies defined in the operators $\hat{\bm{\theta}}$ are:
  \begin{align}
    \omega^u_{kpq} &= \omega ^{u}_{k}+\omega ^{u}_{p}+\omega ^{u}_{q} \qc \omega _{k}^{u}=\nu k^2, \nonumber \\
    \omega^b_{kpq} &= \omega ^{u}_{k}+\omega ^{b}_{p}+\omega ^{b}_{q} \qc \omega _{k}^{b}=\eta k^2 .
    \label{eq:inverse_timescales}
  \end{align}
Isotropy helps to simplify this problem further. By writing the second moment in terms of the rotationally invariant 
second-rank tensors $\delta _{\alpha \beta } $ and $\hat{k}_{\alpha },\hat{k}_{\beta }$:
\begin{equation}
  \zf{XX}_{\alpha \beta }(k)= \qty(c_1\delta _{\alpha \beta }+c_2 \hat{k}_{\alpha }\hat{k}_{\beta }) \\ 
  \label{eq:invariance}
\end{equation}
 and since, by definition, $\zf{XX}_{\alpha \alpha }(k)=\uf{X}(k)$, we obtain for the incompressible problem 
$c_1= \frac{1}{d-1}=-c_2$ 
Incompressibility demands $\hat{k}_{\alpha} \zf{XX}_{\alpha \beta}=0$, hence $c_1+c_2=0$. Introducing the trace of the 
second moment as $\zf{XX}_{\alpha \alpha }(k)=\uf{X}(k)$, allows us to rewrite Eq.~\ref{eq:invariance} as
\begin{equation}
\zf{XX}_{\alpha \beta }(k)= \frac{1}{(d-1)} \, \Pb{k}_{\alpha \beta }\uf{X}(k) .
\label{eq:isotropy}
\end{equation}
Furthermore, the operators within the isotropic model obey:
  \begin{align}
    \Pb{k}_{\alpha \beta }\Pb{k}_{\alpha \rho  }= \Pb{k}_{\beta  \rho }\quad;&\quad \Pb{k}_{\alpha \beta }\Pb{k}_{\alpha \beta }= d-1 \nonumber \\ 
    \Pc{k}_{\alpha \rho \gamma  }= \Pc{k}_{\alpha \gamma \rho }\quad;&\quad \Cc{k}_{\alpha \rho \gamma  }= -\Cc{k}_{\alpha \gamma \rho }\nonumber \\ 
    \Pb{k}_{\alpha \beta } \Pc{k}_{\beta \rho \gamma }= \Pc{k}_{\alpha \rho \gamma }\quad;&\quad  \Pb{k}_{\alpha \beta } \Cc{k}_{\beta \rho \gamma }= \Cc{k}_{\alpha \rho \gamma } %
    \label{eq:proj_n_tran1}
  \end{align}
By exploiting these symmetries, it is then a matter of algebra to show 
\begin{subequations}
\begin{align}
  \dop{2\nu k^2}\uf{u}(k)&= \frac{4}{\qty(d-1)^2} \Itr \Oth{kpq} \Pc{k}_{\alpha \rho \gamma }
  \left[ - \Pc{k}_{\alpha \mu \sigma }\Pb{q}_{\rho \mu }\Pb{p}_{\gamma \sigma }\uf{u}(q)\uf{u}(p) \right. \nonumber \\
  & \hspace{3em} \left. +\Pc{q}_{\rho \mu \sigma }\Pb{k}_{\alpha \sigma }\Pb{p}_{\gamma \mu }\uf{u}(k)\uf{u}(p)+
  \Pc{p}_{\gamma \mu \sigma }\Pb{k}_{\alpha \mu }\Pb{q}_{\rho \sigma }\uf{u}(k)\uf{u}(q) \right] \nonumber  \\
  & -\frac{4}{\qty(d-1)^2} \Itr \OthB{kpq}\Pc{k}_{\alpha \rho \gamma }
  \left[ \Pc{k}_{\alpha \mu \sigma }\Pb{q}_{\rho \mu }\Pb{p}_{\gamma \sigma }\uf{b}(q)\uf{b}(p) \right. \nonumber \\
  & \hspace{3em} \left. +\Cc{q}_{\rho \mu \sigma }\Pb{k}_{\alpha \sigma }\Pb{p}_{\gamma \mu }\uf{u}(k)\uf{b}(p)+
   \Cc{p}_{\gamma \mu \sigma }\Pb{k}_{\alpha \mu }\Pb{q}_{\rho \sigma }\uf{u}(k)\uf{b}(q) \right]  \label{eq:2-order_mom_V_qn} \\
  \dop{2\eta k^2}\uf{b}(k)&= \frac{4}{\qty(d-1)^2}\Itr \OthB{pqk} \Cc{k}_{\alpha \rho \gamma }
  \left[-\Pc{p}_{\gamma \mu \sigma }\Pb{k}_{\alpha \mu }\Pb{q}_{\rho \sigma }\uf{b}(k)\uf{b}(q) \right. \nonumber \\
  & \hspace{3em} \left. -\Cc{k}_{\alpha \mu \sigma } \Pb{q}_{\rho \mu }\Pb{p}_{\gamma \sigma }\uf{b}(q)\uf{u}(p)+\Cc{q}_{\rho \mu \sigma }\Pb{k}_{\alpha \mu }\Pb{p}_{\gamma \sigma }\uf{b}(k)\uf{u}(p) \right] \nonumber \\
& +\frac{4}{\qty(d-1)^2}\Itr \OthB{qkp} \Cc{k}_{\alpha \rho \gamma }
  \left[-\Pc{q}_{\gamma \mu \sigma }\Pb{k}_{\alpha \mu }\Pb{p}_{\rho \sigma }\uf{b}(k)\uf{b}(p) \right. \nonumber \\
  & \hspace{3em} \left. -\Cc{k}_{\alpha \mu \sigma } \Pb{p}_{\rho \mu }\Pb{q}_{\gamma \sigma }\uf{b}(p)\uf{u}(q)+\Cc{p}_{\rho \mu \sigma }\Pb{k}_{\alpha \mu }\Pb{q}_{\gamma \sigma }\uf{b}(k)\uf{u}(q) \right]  \label{eq:2-order_mom_B_qn}%
\end{align}%
\label{eq:2-order_mom_iso}%
\end{subequations}
It is useful to introduce geometric coefficients
\begin{subequations}
\begin{align}
  \gc{a}_{kqp}= \gc{a}_{kpq}&= -\frac{1}{k^2} \Pc{k}_{\alpha \rho \gamma }\Pc{k}_{\alpha \mu \sigma }\Pb{q}_{\rho \mu }\Pb{p}_{\gamma \sigma } \label{eq:geom_a} \\
  \gc{b}_{kpq}&= -\frac{2}{k^2} \Pc{k}_{\mu  \rho \gamma }\Pc{p}_{\gamma \mu \sigma }\Pb{q}_{\rho \sigma } \label{eq:geom_b} \\
  \gc{c}_{kpq}&= \frac{2}{k^2} \Pc{k}_{\mu  \rho \gamma }\Cc{p}_{\gamma \mu \sigma }\Pb{q}_{\rho \sigma } \label{eq:geom_c} \\
  \gc{g}_{kqp}=  \gc{g}_{kpq}&= -\frac{2}{k^2}\Cc{k}_{\alpha \rho \gamma }\Cc{k}_{\alpha \mu \sigma }\Pb{q}_{\rho \mu }\Pb{p}_{\gamma \sigma }  \label{eq:geom_g}\\
  \gc{h}_{kpq}&= -\frac{2}{k^2} \Cc{k}_{\mu  \rho \gamma }\Cc{p}_{\rho \mu  \sigma }\Pb{q}_{\gamma \sigma } \label{eq:geom_h} 
\end{align}
\label{eq:geom_coeff}%
\end{subequations}
These geometric coefficients depend on the angle of the triangle formed by
$\vb{k,p,q}$, and expressed in terms of the cosines of the resultant angles.
Specifically, for a triangle formed by sides of length $k,p,q$ and defining the
cosines of the angles opposite to their sides as $x,y,z$, the following
relations hold~\cite{pouquet_strong_1976, schilling_triadic_2002,
clark_effect_2021}
\begin{subequations}
\begin{align}
x^2+y^2+z^2&=1-2xyz \nonumber \\
\gc{a}_{kpq}&= \frac{1}{2}\qty(1-2y^2z^2-xyz+\dfrac{(d-3)}{2}\qty[2-y^2-z^2]) \label{eq:geom_a_val} \\
\gc{b}_{kpq}&= \dfrac{p}{k}\qty(z^3+xy+\dfrac{(d-3)}{2}\qty[z+xy])\label{eq:geom_b_val}  \\
\gc{c}_{kpq}&= \dfrac{p}{k}\qty(z\qty[1-y^2]+\dfrac{(d-3)}{2}\qty[z+xy])  \label{eq:geom_c_val} \\
\gc{g}_{kpq}&= 1+xyz+\dfrac{(d-3)}{2}\qty[2-y^2-z^2] \label{eq:geom_f_val} \\
\gc{h}_{kpq}&= \dfrac{p}{2k}\qty( (d-1)\qty[z+xy])\label{eq:geom_h_val}
\end{align}
\label{eq:geom_coeff_val}%
\end{subequations}
Formally, the evolution equation for the spectral energies can be written in the form 
\begin{align}
  \dop{2\nu k^2}\uf{X}&= \Itr \tf{X}(\Trd),
  \label{eq:2-order_mom_transfer0}
\end{align}
where $\tf{X}(\Trd)$ is the transfer integrand for the field $X$ arising from a
particular triad $\vb{k,p,q}$. The constructed integrand depends only on the
geometry of the triad; this allows us to integrate out additional degrees of freedom in such
integral $\Itr$. 

\begin{figure}[t!] \centering
  \includegraphics[width=0.4\textwidth]{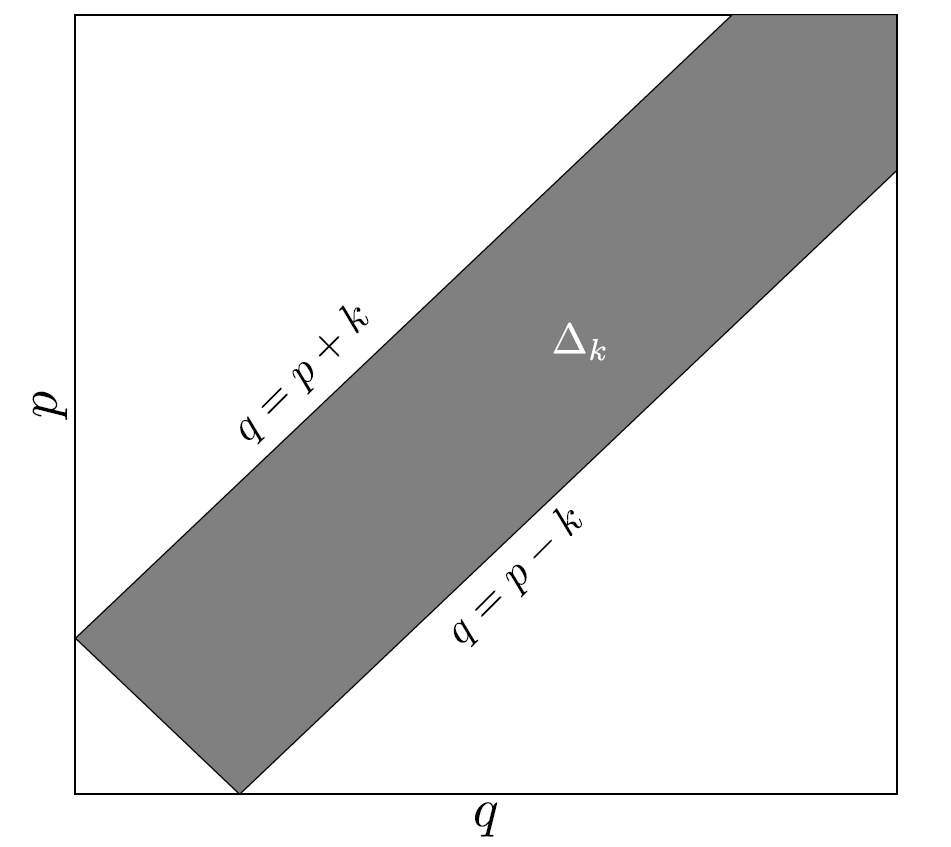}
  \caption[Area of integration for the spectral transfer term]{Showing the area of
    integration in the $p-q$ plane, for a given $k$, that satisfies the
    triangle inequality for the sides $k,p,q$, hence contributing to the
  integral in the transfer term $\tf{X}$.}
  \label{fig:kpq_integration_plane}
\end{figure}

In $d$-dimensions such integrals can be simplified as follows. By construction, 
the transfer term is a function of just the magnitude and angle for a pair 
of wavevectors $\vb{k}$ and $\vb{q}$. In $d$ dimensions, the Cartesian coordinates, radius, and 
spherical angles are related as 
  \begin{align}
    x_1&= r \cos(\phi _1) \nonumber \\ 
    x_2&= r \sin(\phi_1 ) \cos(\phi _2)\nonumber \\ 
    \vdots & \quad \vdots \nonumber \\ 
    x_{n-1}&= r \sin(\phi _1) \cdots \sin(\phi _{n-2}) \cos(\phi _{n-1})\nonumber \\ 
    x_{n}&= r \sin(\phi _1) \cdots \sin(\phi _{n-2}) \sin(\phi _{n-1}) 
    \label{eq:d_dim_co}
  \end{align}
  We align, for convenience, our axis such that $\vb{k}$ is along $x_1$ and denote $\phi _1=\beta$. 
  We now integrate out the remaining angles $\phi _2, \cdots, \phi _{n-1}$ which form a
  $d-2$ dimensional sphere yielding
  \begin{equation}
    \int_{}^{}\dd[d-1]{\Omega } = \mathrm{S}_{d-1}= \dfrac{2 \pi ^{\nicefrac{d}{2} }}{\Gamma (\frac{d}{2})}. 
    \label{eq:d_surface}
  \end{equation}
This now allows us to further evaluate the evolution equation for the spectral energies as:
\begin{align}
  \dop{2\nu k^2}\uf{X}&= \Itr \tf{X}(\Trd) \nonumber \\
  &= \displaystyle \int_{}^{}\dd{\vb{q}} \dd{\vb{p}}\bm{\delta }\qty(\vb{p}+\vb{q}-\vb{k}) \tf{X}(\Trd) \nonumber \\ 
  &= \displaystyle \int_{}^{}\dd[d]{q} \tf{X}(\Trd)= \displaystyle \int_{}^{}\dd{q} q^{d-1} \dd[d-1]{\Omega }\tf{X}(\Trd)\nonumber \\ 
  &= \displaystyle \int_{q=0}^{\infty} \dd{q}\; q^{d-1} \displaystyle \int_{\beta =0}^{\pi }\dd{\beta}\; \sin^{d-2}(\beta )\tf{X}(\Trd) \displaystyle \int_{}^{}\dd[d-2]{\Omega } \nonumber \\ 
  &= \displaystyle \int_{q=0}^{\infty} \dd{q}\; q^{d-1} \displaystyle \int_{y=-1}^{y=1 }\dd{y}\; \sin^{d-3}(\beta )\tf{X}(\Trd) \mathrm{S}_{d-2}  
\label{eq:2-order_mom_transfer1}
\end{align}
  To further simplify Eq.~\eqref{eq:2-order_mom_transfer1}, we use the sine law of triangle and following change of variables:
  \begin{align}
    \frac{\sin(\alpha )}{k}&= \frac{\sin(\beta )}{p} \nonumber \\ 
  \displaystyle \int_{0}^{\infty}\dd{q}\; \displaystyle \int_{0}^{1} \dd{y}\; &=  \vb{J}\qty[\dfrac{\partial \qty( q(q,p),y(q,p))}{\partial \qty(q,p)}] \displaystyle \int_{0}^{\infty}\dd{q} \; \displaystyle \int_{\abs{k-q}}^{\abs{k+q}}\dd{p}  \nonumber \\ 
    &= \qty(\frac{p}{kq}) \Iptr \DqDp 
    \label{eq:triad_integration}
  \end{align}
  where $\vb{J}\qty[\dfrac{\partial \qty(q(q,p),y(q,p))}{\partial \qty(q,p)}]$
  is the Jacobian for change of variables, and $\Iptr \DqDp=\displaystyle
  \int_{0}^{\infty}\dd{q} \; \displaystyle \int_{\abs{k-q}}^{\abs{k+q}}\dd{p}
  $.  By using the relation \eqref{eq:triad_integration} in Eq.~\eqref{eq:2-order_mom_transfer1}, we finally arrive at 
\begin{align}
  \dop{2\nu k^2}\uf{X}&= \mathrm{S}_{d-2} \Iptr \DqDp \qty(\dfrac{pq}{k})^{d-2} \qty(1-x^2)^{\tfrac{(d-3)}{2} } \tf{X}(\Trd).
  \label{eq:2-order_mom_transfer2}
\end{align}
Here $\mathrm{S}_{d}$ is the solid angle of a $d-$dimensional sphere. Now
we have to integrate Eq.~\eqref{eq:2-order_mom_transfer2} over the $p-q$
plane that can form a triangle with a side of length $k=\abs{\vb{k}}$. By
using the triangle inequality, this region would involve $p+q<k<\abs{p-q}$, as shown in
Fig.~\ref{fig:kpq_integration_plane}, for every $k$.
Isotropy implies that the energy spectrum and spectral energy are related by
\begin{align}
  \uf{X}(k)&= \dfrac{2\ef{X}(k)}{k^{d-1}\mathrm{S}_{d-1}}
  \label{eq:spectrum_relation}
\end{align}
All the geometric coefficients are not independent. In fact it is not difficult to prove the 
following constraints:
\begin{subequations}
\begin{align}
  2\gc{a}_{kpq}&= \gc{b}_{kpq} + \gc{b}_{kqp} \label{eq:geom_I_ab}\\
  2\gc{a}_{kpq}&= \gc{c}_{kpq} + \gc{c}_{kqp} \label{eq:geom_I_ac}\\
  k^2 \gc{b}_{kpq}&= p^2 \gc{b}_{pkq} \label{eq:geom_I_b}\\
  k^2 \gc{h}_{kpq}&= p^2 \gc{h}_{pkq} \label{eq:geom_I_h}\\
  \gc{g}_{kpq}&= \gc{h}_{kpq}+\dfrac{q^2}{k^2}\gc{c}_{qkp} \label{eq:geom_I_ghc}\\
  q^4\gcr{c}_{qkp}&= k^4\gcr{c}_{kqp} \label{eq:geom_I_c} \\
  \qty(p^2 \gc{c}_{kqp}+q^2 \gc{c}_{kpq})&= \qty(p^2 \gc{h}_{kqp}+q^2 \gc{h}_{kpq}) \dfrac{(d-2)}{(d-1)} \label{eq:geom_I_ch} \\
  k^2 \gcw{a}_{kpq} &= p^2 \gcw{b}_{kpq}+q^2 \gcw{b}_{kqp} \label{eq:geom_I_b2} \\
  k^2 \gcw{c}_{kqp}&= q^2 \qty( \gcw{h}_{pqk}-\gcw{h}_{qpk}) \label{eq:geom_I_c2}
\end{align}
  \label{eq:geom_I}%
\end{subequations}
With the superscripts $(2)$ and $(3)$ corresponding to two and three dimensions, respectively. Finally, exploiting 
the symmetry between $p$ and $q$, we construct the quasi-Normal MHD equations: 
\begin{subequations}
  \begin{align}
    \dop{2\nu k^2}\ef{u}(k,t)&= 8K_d \Iptr \DqDp \qty[\Btrf{u}{uu}\qty(k,p,q)+\Btrf{u}{bb}\qty(k,p,q)+\Btrf{u}{ub}\qty(k,p,q)]\; W_d\qty(\Trd)  \label{eq:qn_mhd_V2}\\
    \dop{2\eta  k^2}\ef{b}(k,t)&= 8K_d \Iptr \DqDp \qty[\Btrf{b}{ub}\qty(k,p,q)+\Btrf{b}{bb}\qty(k,p,q)]\; W_d\qty(\Trd)  \label{eq:qn_mhd_B2}\\
    \Btrf{u}{uu}\qty(k,p,q)&= \Oth{kpq}\frac{k}{pq}\gc{b}_{kpq}\qty[k^{d-1}\ef{u}(p)-p^{d-1}\ef{u}(k)]\ef{u}(q)  \label{eq:qn_mhd_tvvv2}\\
    \Btrf{u}{bb}\qty(k,p,q)&= \OthB{kpq}\frac{k}{pq}\gc{c}_{kpq} k^{d-1}\ef{b}(p)\ef{b}(q)  \label{eq:qn_mhd_tvbb2}\\
    \Btrf{u}{ub}\qty(k,p,q)&= -\OthB{kpq}\frac{k}{pq}\gc{c}_{kpq} p^{d-1}\ef{u}(k)\ef{b}(q) \label{eq:qn_mhd_tvvb2}\\
    \Btrf{b}{ub}\qty(k,p,q)&= \OthB{qkp}\frac{k}{pq}\gc{h}_{kpq}\qty[k^{d-1}\ef{b}(p)-p^{d-1}\ef{b}(k)] \ef{u}(q) 
  + \OthB{pqk}\frac{p}{kq}\gc{c}_{pkq}k^{d-1}\ef{u}(p)\ef{b}(q) \label{eq:qn_mhd_tbvb2}\\
  \Btrf{b}{bb}\qty(k,p,q)&= -\OthB{pqk} \frac{p}{kq}\gc{c}_{pkq}p^{d-1} \ef{b}(k)\ef{b}(q) \label{eq:qn_mhd_tbbb2}
  \end{align}
  \label{eq:qn_mhd}
\end{subequations}
The dimensional pre-factor $K_d$ and the triad weights $W_d(\Trd)$ are 
\begin{subequations}
\begin{align}
K_d&=\frac{1}{\qty(d-1)^{2}}\dfrac{\mathrm{S}_{d-2}}{\mathrm{S}_{d-1}} \label{eq:K_d} \\
W_d(\Trd)& = \qty(\dfrac{\sin^2 x}{k^2})^{\nicefrac{(d-3)}{2} }
\label{eq:W_d}
\end{align}
\label{eq:prefactors}%
\end{subequations}
This is a closed set of integro-differential equations, that can be solved
numerically. Note that the  operators $\Oth{kpq}$ and $\OthB{kpq}$ involve time integrals, as
defined in Eq.~\eqref{eq:def_inverse}. This QN model (Eqs.~\eqref{eq:qn_mhd})
respects the conservation laws that the original \pde has. The sum of kinetic
and magnetic energy is conserved for the ideal fluid, that is $\nu =\eta =0$.
Particularly, in two-dimensions, the net magnetic potential is conserved and
for a pure kinetic model ($\vb{B}=0$) the enstrophy remains conserved. 

\section{Appendix B: The Eddy-Damped Quasi-Normal Approximation} 
\renewcommand{\theequation}{B-\arabic{equation}}
\setcounter{equation}{0}  % reset counter 

We now make two further approximations in the spirit of the \edq
model~\cite{orszag_analytical_1970, lesieur_turbulence_2008, sagaut_edqnm_2002,
sagaut_homogeneous_2018} for hydrodynamic turbulence. This is because quasi Normal 
by itself does not ensure the positive definite nature of kinetic energy owing to 
divergences in the third-order moments. This is cured by \textit{damping}, which ensures 
a saturation of the third-order moments by introducing an inverse time-scale $\mu _{kpq}$ 
for the triad $k,p,q$,
\begin{align}
  \dop{\nu \qty(k^2+p^2+q^2)+\mu _{kpq}}\ev{uuu} &=\ev{uu} \ev{uu},  \nonumber \\
  \mu _{kpq}&= \mu _k + \mu _p+ \mu _q,
  \label{eq:eddy_damping_def1}
\end{align}
which ought to originate from the spectrum.
On dimensional grounds this is simply
\begin{align}
  \mu _k &\approx \qty[k^3\ef{u}(k)]^{\frac{1}{2}}.
  \label{eq:eddy_damping_def2}
\end{align}
Further improvement of this
\begin{align}
  \mu _k &= \alpha_d \qty[\displaystyle \int_{0}^{k}\dd{p} p^2 \ef{u}(p)]^{\frac{1}{2}}
  \label{eq:eddy_damping_def3}
\end{align}
factors in the deformation of eddies of size $k^{-1}$  by  larger  eddies.  This
allows for a free parameter $\alpha_d$ which, for fluid  turbulence,  fixes  the
$d$-dimensional Kolmogorov constant $C_d$ for the corresponding  $d$-dimensional
kinetic energy spectrum $E(k)=C_dk^{-5/3}\epsilon^{2/3}$. For a  given  $d$,
$C_d$    is    chosen    from    interpolating    the    values     given     in
Ref.~\cite{clark_effect_2021}.

We use the same inspiration, for our closure model, to define (for the magnetic field) 
\begin{equation}
  \mu_k =\alpha_d  \qty[\displaystyle \int_{0}^{k}\dd{p} p^2 \ef{b}(p)]^{\frac{1}{2}} + \sqrt{\dfrac{2}{3}} \qty(\displaystyle \int_{0}^{k}\dd{ p} \ef{b}(p))^{\frac{1}{2}}
  \label{eq:eddy_damping_alfven}
\end{equation}
with  the  additional  piece  accounting  for  the  effects  of   the   Alfv\'en
waves~\cite{davidson_introduction_2001}; the coefficient $\sqrt{2/3}$ comes from
an explicit calculation of the Alfv\'{e}n timescales for a Gaussian  large-scale
magnetic fields~\cite{pouquet_strong_1976}.

In summary, the final eddy-damping time-scales, acting linearly  on  third-order
moments in our closure model, are given by:
\begin{align}
\omega _{k}^u&= \nu k^2 + \alpha_d \qty[\displaystyle \int_{0}^{k}\dd{p} p^2 \ef{V}(p)]^{\frac{1}{2}}; \nonumber \\
\omega _{k}^b&= \eta k^2 + \alpha_d  \qty[\displaystyle \int_{0}^{k}\dd{p} p^2 \ef{b}(p)]^{\frac{1}{2}} + \sqrt{\dfrac{2}{3}} \qty(\displaystyle \int_{0}^{k}\dd{ p} \ef{b}(p))^{\frac{1}{2}}.
\label{eq:eddy_timescales}
\end{align}

\section{Appendix C: The Eddy-Damped Quasi-Normal Markovian Model} 
\renewcommand{\theequation}{C-\arabic{equation}}
\setcounter{equation}{0}  % reset counter 

The  eddy-damping  time-scales  by  itself  does  not  guarantee  the   positive
definiteness  of  the  the  energy  spectrum.  A  final  approximation,  due  to
Orszag~~\cite{orszag_analytical_1970}, is \emph{Markovianization}. This  assumes
that the third-order moments vary slowly when compared  to  the  exponential
decay in the  $\Oth{kpq}$ operator. This separation of time-scales  allows
the approximation, where the time integral in $\Oth{kpq}$  is
computed explicitly, leading to further simplification
\begin{equation}
  \Oth{kpq}=  \displaystyle  \int_{0}^{t}\dd{s}   \exp(-\omega^u_{kpq}(t-s))   = \\dfrac{1-\exp(-\omega^u_{kpq}t)}{\omega^u_{kpq}}. label{eq:markovianization}
\end{equation}
Further,  in  the  large  $t$   limit,   $\Oth{kpq}=   1/\omega_{kpq}^u   \equiv
\theta^{u}_{kpq}$ and $\OthB{kpq}=1/\omega  _{kpq}^{b}\equiv  \theta^{b}_{kpq}$.
This establishes an instantaneous relationship between third  and  second  order
moments and makes the process memoryless or Markovian.

We are now in a position, with all our ingredients in place, to write the final set of 
$d$-dimensional, MHD-EDQNM equations for incompressible, magnetohydrodynamic turbulence:
\begin{subequations}
\begin{align}
\dv{}{t}\ef{u}(k,t)&= \tnf{u}(k,t)- 2\nu k^2 \ef{u}(k,t); \label{eq:edqnm-mhd_spectrum_u}\\
\dv{}{t}\ef{b}(k,t)&= \tnf{b}(k,t)- 2\eta k^2 \ef{b}(k,t) \label{eq:edqnm-mhd_spectrum_b}.
\end{align}
\label{eq:edqnm-mhd_spectrum}%
\end{subequations}
It is extremely useful to study how the kinetic energy and
magnetic energy spectrum interacts in the transfer terms.  In order to do that,
divide the contributions to the transfer terms, as done in the main text, as \emph{self} (subscript (s))  and
\emph{coupled} (subscript (c)):
\begin{subequations}
\begin{align}
\tnf{u}(k)&= \tn{u}{(s)}(k)+\tn{u}{(c)}(k); \label{eq:transfer_split_u}\\
\tnf{b}(k)&= \tn{b}{(s)}(k)+\tn{b}{(c)}(k); \label{eq:transfer_split_b}
\end{align}
\label{eq:transfer_split}
\end{subequations}
where
\begin{subequations}
\begin{align}
\tn{u}{(s)}(k)&= 8K_d \Iptr \DqDp W_d\qty(\Trd) \theta^u_{kpq}\frac{k}{pq}\gc{b}_{kpq}\qty[k^{d-1}\ef{u}(p)-p^{d-1}\ef{u}(k)]\ef{u}(q) \label{eq:transfer_u_s}\\
\tn{u}{(c)}(k)&= 8K_d \Iptr \DqDp W_d\qty(\Trd) \theta^b_{kpq}\frac{k}{pq}\gc{c}_{kpq}\qty[k^{d-1}\ef{b}(p)-p^{d-1}\ef{u}(k)]\ef{b}(q) \label{eq:transfer_u_c}\\
\tn{b}{(s)}(k)&= 8K_d \Iptr \DqDp W_d\qty(\Trd) \theta^b_{qkp}\frac{k}{pq}\gc{h}_{kpq}\qty[k^{d-1}\ef{b}(p)-p^{d-1}\ef{b}(k)]\ef{u}(q) \label{eq:transfer_B_s}\\
\tn{b}{(c)}(k)&= 8K_d \Iptr \DqDp W_d\qty(\Trd) \theta^b_{pqk}\frac{p}{kq}\gc{c}_{pkq}\qty[k^{d-1}\ef{u}(p)-p^{d-1}\ef{b}(k)]\ef{b}(q) \label{eq:transfer_B_c}
\end{align}
\label{eq:transfer_split2}
\end{subequations}

\section{Appendix D: Numerical Simulations of the $d$-dimensional MHD-EDQNM Model} 
\renewcommand{\theequation}{D-\arabic{equation}}
\setcounter{equation}{0}  % reset counter 

Given that the original \edqmhd equations have infinite degrees of freedom, to
numerically study them we have to discretize the wavenumber space, say
$\mathcal{D}$. Since we are expecting a power-law behavior for the energy spectrum
in the inertial range, and want to achieve high Reynolds numbers (both kinetic
and magnetic) in the simulation, it is easier if we discretise the $N$ wavenumbers $\qty{k_i}$
in a geometric sequence as follows:
\begin{equation}
\mathcal{D}\equiv \{k_i=k_1\lambda ^{i-1} \},\; i=1,2, \cdots , N.
  \label{eq:wavenumber_domain}
\end{equation}
The wavenumber bands are chosen as $\Delta k_i=k_i \ln
\lambda $.  Suppose we denote the upper $k_i^+ $and lower  $k_i^- $ limits of
the $i^{\mathrm{th}}$ band by
\begin{subequations}
\begin{align}
	k_{i}^{+}&= k_i+\Delta k_{i}^{+} \\
	k_{i}^{-}&= k_i-\Delta k_{i}^{-} \\
	\Delta k_{i}^{+}+\Delta k_{i}^{-}&= \Delta k_{i}^{}
\end{align}
  \label{eq:wavenumber_domain2}%
\end{subequations}
Since we wish to cover the whole of the wavenumber space till
$k_N$ without any gaps or overlaps, the lower limit of the
$(i+1)^{\mathrm{th}}$ band should coincide with the upper limit of the
$i^{\mathrm{th}}$ band:
\begin{subequations}
\begin{align}
	k_i+\Delta k_{i}^{+}&= k_{i+1}-\Delta k_{i+1}^{-}\\
	\Delta k_{i+1}^{-}&= \Delta k_{i}^{-}+k_{i+1}\left( 1-\frac{1}{\lambda }-\frac{\ln \lambda }{\lambda } \right)\\
	&= \Delta k_{i}^{-}+k_1\left( 1-\frac{1}{\lambda }-\frac{\ln \lambda }{\lambda } \right)\dfrac{\lambda ^i-\lambda }{\lambda -1}\\
	k_{i}^{+}&= \frac{\Delta k_i}{\lambda -1}+\frac{k_1}{\lambda -1}\left( \lambda -1-\ln \lambda  \right)-\Delta k_{1}^{-}\\
	k_{i}^{-}&= \frac{\Delta k_{i-1}}{\lambda -1}+\frac{k_1}{\lambda -1}\left( \lambda -1-\ln \lambda  \right)-\Delta k_{1}^{-}
\end{align}
  \label{eq:wavenumber_domain3}%
\end{subequations}
Now, without loss of generality, we choose $\Delta
k_{1}^{-}=k_1 \dfrac{\left( \lambda -1-\ln \lambda  \right)}{\lambda -1}$, whence
\begin{subequations}
\begin{align}
	k_{i}^{-}&= \frac{\Delta k_i}{\lambda -1}\\
	k_{i}^{+}&= \frac{\Delta k_{i+1}}{\lambda -1}
\end{align}
  \label{eq:wavenumber_domain4}%
\end{subequations}
In this framework of discrete wavenumber bands, the integral in the transfer terms becomes:
\begin{subequations}
\begin{align}
  \Iptr \DqDp \Big|_{k=k_i} & \equiv \displaystyle \sum_{j=1}^{N}\:\displaystyle \sum_{l=l_{\mathrm{min}}}^{l_{\mathrm{max}}} \\
	l_{\mathrm{min}}(i,j)&=\left[  \mathrm{Log}_{\lambda } \left( |k_i-k_j| \right) \right]_{>} +1\\
	l_{\mathrm{max}}(i,j)&= \mathrm{Max}\left\{ N, \left[ \mathrm{Log}_{\lambda } \left( k_i+k_j \right) \right]_{<}+1 \right\}
\end{align}
  \label{eq:code_parameters}%
\end{subequations}
The integration limits are chosen such that $k_i,k_j,k_l
\in \mathcal{D} $ can form a triangle. In the above equations, $[x]_<$ and $[x]_>$ correspond to the lowest and the
greatest integer function.

%%%%%%%%%%%%%%%%%%%%%%%%%%%%%%%%
\end{document}